\pdfoutput=1
\documentclass[aps,prd,amsmath,amssymb,superscriptaddress, nofootinbib, longbibliography, showkeys, reprint,preprintnumbers]{revtex4-2}

\usepackage{color}
\usepackage{aas_macros}
\usepackage{hyperref,breakurl}
\hypersetup{colorlinks=true, citecolor=blue, urlcolor=blue, linkcolor=blue}
\usepackage[usenames,dvipsnames]{xcolor}
\usepackage{dcolumn}
\usepackage{bm}
\usepackage{array}
\usepackage{graphicx}
\usepackage{graphics}
\usepackage{enumitem}
\usepackage{xspace}
\usepackage{amsmath,amssymb,latexsym,times}

\usepackage{float}

\usepackage{todonotes}
\usepackage{changes} 
\usepackage[normalem]{ulem} 

\newcommand{\der}{\mathrm{d}}

\newcommand{\beq}{\begin{equation}}
\newcommand{\eeq}{\end{equation}}

\newcommand{\code}[1]{\texttt{#1}\xspace}

\newcommand{\cosmolike}{\textsc{CosmoLike}}
\newcommand{\cosmosis}{\textsc{CosmoSIS}}

\newcommand{\dnf}{\code{DNF}}
\newcommand{\bpz}{\code{BPZ}}
\newcommand{\gold}{\code{Y3 GOLD}}

\newcommand{\sof}{SOF\xspace}
\newcommand{\metacal}{\textsc{Metacalibration}\xspace}
\newcommand{\redmapper}{\textsc{redMaPPer}}
\newcommand{\redmagic}{\textsc{redMaGiC}\xspace}
\newcommand{\maglim}{\textsc{MagLim}\xspace}
\newcommand{\sextractor}{\code{SExtractor}\xspace}

\newcommand{\photoz}{photo-$z$\xspace}

\newcommand{\delg}{\delta_{\rm g}}

\newcommand{\LCDM}{$ \Lambda $CDM}

\newcommand{\omb}{\Omega_{\rm b}}

\newcommand{\Mpc}{\mathrm{Mpc}}

\newcommand{\sqdeg}{{\rm deg}^{2}}

\newcommand{\nside}{\ifmmode N_{\mathrm{side}}\else $N_{\mathrm{side}}$\fi}
\newcommand{\npix}{\ifmmode n_{\mathrm{pix}}\else $n_{\mathrm{pix}}$\fi}
\newcommand{\Npix}{\ifmmode N_{\mathrm{pix}}\else $n_{\mathrm{pix}}$\fi}
\newcommand{\lmin}{\ifmmode \ell_{\mathrm{min}}\else $\ell_{\mathrm{min}}$\fi}
\newcommand{\lmax}{\ifmmode \ell_{\mathrm{max}}\else $\ell_{\mathrm{max}}$\fi}

\newcommand\Tstrut{\rule{0pt}{2.6ex}}         
\newcommand\Bstrut{\rule[-0.9ex]{0pt}{0pt}}   

\definecolor{wine}{RGB}{153,60,51}

\definecolor{blue-violet}{rgb}{0.42, 0.17, 0.89}

\definechangesauthor[color=blue-violet]{AP}
\setauthormarkuptext{}

\defcitealias{Tanoglidis2019}{T20}

\begin{document}

\ \
\title{Dark Energy Survey Year 3 results: Optimizing the lens sample in a combined galaxy clustering and galaxy-galaxy lensing analysis}


\author{A.~Porredon}
\affiliation{Center for Cosmology and Astro-Particle Physics, The Ohio State University, Columbus, OH 43210, USA}
\affiliation{Department of Physics, The Ohio State University, Columbus, OH 43210, USA}
\affiliation{Institut d'Estudis Espacials de Catalunya (IEEC), 08034 Barcelona, Spain}
\affiliation{Institute of Space Sciences (ICE, CSIC),  Campus UAB, Carrer de Can Magrans, s/n,  08193 Barcelona, Spain}
\author{M.~Crocce}
\affiliation{Institut d'Estudis Espacials de Catalunya (IEEC), 08034 Barcelona, Spain}
\affiliation{Institute of Space Sciences (ICE, CSIC),  Campus UAB, Carrer de Can Magrans, s/n,  08193 Barcelona, Spain}
\author{P.~Fosalba}
\affiliation{Institut d'Estudis Espacials de Catalunya (IEEC), 08034 Barcelona, Spain}
\affiliation{Institute of Space Sciences (ICE, CSIC),  Campus UAB, Carrer de Can Magrans, s/n,  08193 Barcelona, Spain}
\author{J.~Elvin-Poole}
\affiliation{Center for Cosmology and Astro-Particle Physics, The Ohio State University, Columbus, OH 43210, USA}
\affiliation{Department of Physics, The Ohio State University, Columbus, OH 43210, USA}
\author{A.~Carnero~Rosell}
\affiliation{Instituto de Astrofisica de Canarias, E-38205 La Laguna, Tenerife, Spain}
\affiliation{Universidad de La Laguna, Dpto. Astrofísica, E-38206 La Laguna, Tenerife, Spain}
\author{R.~Cawthon}
\affiliation{Physics Department, 2320 Chamberlin Hall, University of Wisconsin-Madison, 1150 University Avenue Madison, WI  53706-1390}
\author{T.~F.~Eifler}
\affiliation{Department of Astronomy/Steward Observatory, University of Arizona, 933 North Cherry Avenue, Tucson, AZ 85721-0065, USA}
\affiliation{Jet Propulsion Laboratory, California Institute of Technology, 4800 Oak Grove Dr., Pasadena, CA 91109, USA}
\author{X.~Fang}
\affiliation{Department of Astronomy/Steward Observatory, University of Arizona, 933 North Cherry Avenue, Tucson, AZ 85721-0065, USA}
\author{I.~Ferrero}
\affiliation{Institute of Theoretical Astrophysics, University of Oslo. P.O. Box 1029 Blindern, NO-0315 Oslo, Norway}
\author{E.~Krause}
\affiliation{Department of Astronomy/Steward Observatory, University of Arizona, 933 North Cherry Avenue, Tucson, AZ 85721-0065, USA}
\author{N.~MacCrann}
\affiliation{Department of Applied Mathematics and Theoretical Physics, University of Cambridge, Cambridge CB3 0WA, UK}
\author{N.~Weaverdyck}
\affiliation{Department of Physics, University of Michigan, Ann Arbor, MI 48109, USA}
\author{T.~M.~C.~Abbott}
\affiliation{Cerro Tololo Inter-American Observatory, NSF's National Optical-Infrared Astronomy Research Laboratory, Casilla 603, La Serena, Chile}
\author{M.~Aguena}
\affiliation{Departamento de F\'isica Matem\'atica, Instituto de F\'isica, Universidade de S\~ao Paulo, CP 66318, S\~ao Paulo, SP, 05314-970, Brazil}
\affiliation{Laborat\'orio Interinstitucional de e-Astronomia - LIneA, Rua Gal. Jos\'e Cristino 77, Rio de Janeiro, RJ - 20921-400, Brazil}
\author{S.~Allam}
\affiliation{Fermi National Accelerator Laboratory, P. O. Box 500, Batavia, IL 60510, USA}
\author{A.~Amon}
\affiliation{Kavli Institute for Particle Astrophysics \& Cosmology, P. O. Box 2450, Stanford University, Stanford, CA 94305, USA}
\author{S.~Avila}
\affiliation{Instituto de Fisica Teorica UAM/CSIC, Universidad Autonoma de Madrid, 28049 Madrid, Spain}
\author{D.~Bacon}
\affiliation{Institute of Cosmology and Gravitation, University of Portsmouth, Portsmouth, PO1 3FX, UK}
\author{E.~Bertin}
\affiliation{CNRS, UMR 7095, Institut d'Astrophysique de Paris, F-75014, Paris, France}
\affiliation{Sorbonne Universit\'es, UPMC Univ Paris 06, UMR 7095, Institut d'Astrophysique de Paris, F-75014, Paris, France}
\author{S.~Bhargava}
\affiliation{Department of Physics and Astronomy, Pevensey Building, University of Sussex, Brighton, BN1 9QH, UK}
\author{S.~L.~Bridle}
\affiliation{Jodrell Bank Center for Astrophysics, School of Physics and Astronomy, University of Manchester, Oxford Road, Manchester, M13 9PL, UK}
\author{D.~Brooks}
\affiliation{Department of Physics \& Astronomy, University College London, Gower Street, London, WC1E 6BT, UK}
\author{M.~Carrasco~Kind}
\affiliation{Department of Astronomy, University of Illinois at Urbana-Champaign, 1002 W. Green Street, Urbana, IL 61801, USA}
\affiliation{National Center for Supercomputing Applications, 1205 West Clark St., Urbana, IL 61801, USA}
\author{J.~Carretero}
\affiliation{Institut de F\'{\i}sica d'Altes Energies (IFAE), The Barcelona Institute of Science and Technology, Campus UAB, 08193 Bellaterra (Barcelona) Spain}
\author{F.~J.~Castander}
\affiliation{Institut d'Estudis Espacials de Catalunya (IEEC), 08034 Barcelona, Spain}
\affiliation{Institute of Space Sciences (ICE, CSIC),  Campus UAB, Carrer de Can Magrans, s/n,  08193 Barcelona, Spain}
\author{A.~Choi}
\affiliation{Center for Cosmology and Astro-Particle Physics, The Ohio State University, Columbus, OH 43210, USA}
\author{M.~Costanzi}
\affiliation{INAF-Osservatorio Astronomico di Trieste, via G. B. Tiepolo 11, I-34143 Trieste, Italy}
\affiliation{Institute for Fundamental Physics of the Universe, Via Beirut 2, 34014 Trieste, Italy}
\author{L.~N.~da Costa}
\affiliation{Laborat\'orio Interinstitucional de e-Astronomia - LIneA, Rua Gal. Jos\'e Cristino 77, Rio de Janeiro, RJ - 20921-400, Brazil}
\affiliation{Observat\'orio Nacional, Rua Gal. Jos\'e Cristino 77, Rio de Janeiro, RJ - 20921-400, Brazil}
\author{M.~E.~S.~Pereira}
\affiliation{Department of Physics, University of Michigan, Ann Arbor, MI 48109, USA}
\author{J.~De~Vicente}
\affiliation{Centro de Investigaciones Energ\'eticas, Medioambientales y Tecnol\'ogicas (CIEMAT), Madrid, Spain}
\author{S.~Desai}
\affiliation{Department of Physics, IIT Hyderabad, Kandi, Telangana 502285, India}
\author{H.~T.~Diehl}
\affiliation{Fermi National Accelerator Laboratory, P. O. Box 500, Batavia, IL 60510, USA}
\author{P.~Doel}
\affiliation{Department of Physics \& Astronomy, University College London, Gower Street, London, WC1E 6BT, UK}
\author{A.~Drlica-Wagner}
\affiliation{Department of Astronomy and Astrophysics, University of Chicago, Chicago, IL 60637, USA}
\affiliation{Fermi National Accelerator Laboratory, P. O. Box 500, Batavia, IL 60510, USA}
\affiliation{Kavli Institute for Cosmological Physics, University of Chicago, Chicago, IL 60637, USA}
\author{K.~Eckert}
\affiliation{Department of Physics and Astronomy, University of Pennsylvania, Philadelphia, PA 19104, USA}
\author{A.~Fert\'e}
\affiliation{Jet Propulsion Laboratory, California Institute of Technology, 4800 Oak Grove Dr., Pasadena, CA 91109, USA}
\author{B.~Flaugher}
\affiliation{Fermi National Accelerator Laboratory, P. O. Box 500, Batavia, IL 60510, USA}
\author{J.~Frieman}
\affiliation{Fermi National Accelerator Laboratory, P. O. Box 500, Batavia, IL 60510, USA}
\affiliation{Kavli Institute for Cosmological Physics, University of Chicago, Chicago, IL 60637, USA}
\author{J.~Garc\'ia-Bellido}
\affiliation{Instituto de Fisica Teorica UAM/CSIC, Universidad Autonoma de Madrid, 28049 Madrid, Spain}
\author{E.~Gaztanaga}
\affiliation{Institut d'Estudis Espacials de Catalunya (IEEC), 08034 Barcelona, Spain}
\affiliation{Institute of Space Sciences (ICE, CSIC),  Campus UAB, Carrer de Can Magrans, s/n,  08193 Barcelona, Spain}
\author{D.~W.~Gerdes}
\affiliation{Department of Astronomy, University of Michigan, Ann Arbor, MI 48109, USA}
\affiliation{Department of Physics, University of Michigan, Ann Arbor, MI 48109, USA}
\author{T.~Giannantonio}
\affiliation{Institute of Astronomy, University of Cambridge, Madingley Road, Cambridge CB3 0HA, UK}
\affiliation{Kavli Institute for Cosmology, University of Cambridge, Madingley Road, Cambridge CB3 0HA, UK}
\author{D.~Gruen}
\affiliation{Department of Physics, Stanford University, 382 Via Pueblo Mall, Stanford, CA 94305, USA}
\affiliation{Kavli Institute for Particle Astrophysics \& Cosmology, P. O. Box 2450, Stanford University, Stanford, CA 94305, USA}
\affiliation{SLAC National Accelerator Laboratory, Menlo Park, CA 94025, USA}
\author{R.~A.~Gruendl}
\affiliation{Department of Astronomy, University of Illinois at Urbana-Champaign, 1002 W. Green Street, Urbana, IL 61801, USA}
\affiliation{National Center for Supercomputing Applications, 1205 West Clark St., Urbana, IL 61801, USA}
\author{J.~Gschwend}
\affiliation{Laborat\'orio Interinstitucional de e-Astronomia - LIneA, Rua Gal. Jos\'e Cristino 77, Rio de Janeiro, RJ - 20921-400, Brazil}
\affiliation{Observat\'orio Nacional, Rua Gal. Jos\'e Cristino 77, Rio de Janeiro, RJ - 20921-400, Brazil}
\author{G.~Gutierrez}
\affiliation{Fermi National Accelerator Laboratory, P. O. Box 500, Batavia, IL 60510, USA}
\author{W.~G.~Hartley}
\affiliation{D\'{e}partement de Physique Th\'{e}orique and Center for Astroparticle Physics, Universit\'{e} de Gen\`{e}ve, 24 quai Ernest Ansermet, CH-1211 Geneva, Switzerland}
\author{S.~R.~Hinton}
\affiliation{School of Mathematics and Physics, University of Queensland,  Brisbane, QLD 4072, Australia}
\author{D.~L.~Hollowood}
\affiliation{Santa Cruz Institute for Particle Physics, Santa Cruz, CA 95064, USA}
\author{K.~Honscheid}
\affiliation{Center for Cosmology and Astro-Particle Physics, The Ohio State University, Columbus, OH 43210, USA}
\affiliation{Department of Physics, The Ohio State University, Columbus, OH 43210, USA}
\author{B.~Hoyle}
\affiliation{Faculty of Physics, Ludwig-Maximilians-Universit\"at, Scheinerstr. 1, 81679 Munich, Germany}
\affiliation{Max Planck Institute for Extraterrestrial Physics, Giessenbachstrasse, 85748 Garching, Germany}
\affiliation{Universit\"ats-Sternwarte, Fakult\"at f\"ur Physik, Ludwig-Maximilians Universit\"at M\"unchen, Scheinerstr. 1, 81679 M\"unchen, Germany}
\author{D.~J.~James}
\affiliation{Center for Astrophysics $\vert$ Harvard \& Smithsonian, 60 Garden Street, Cambridge, MA 02138, USA}
\author{M.~Jarvis}
\affiliation{Department of Physics and Astronomy, University of Pennsylvania, Philadelphia, PA 19104, USA}
\author{K.~Kuehn}
\affiliation{Australian Astronomical Optics, Macquarie University, North Ryde, NSW 2113, Australia}
\affiliation{Lowell Observatory, 1400 Mars Hill Rd, Flagstaff, AZ 86001, USA}
\author{N.~Kuropatkin}
\affiliation{Fermi National Accelerator Laboratory, P. O. Box 500, Batavia, IL 60510, USA}
\author{M.~A.~G.~Maia}
\affiliation{Laborat\'orio Interinstitucional de e-Astronomia - LIneA, Rua Gal. Jos\'e Cristino 77, Rio de Janeiro, RJ - 20921-400, Brazil}
\affiliation{Observat\'orio Nacional, Rua Gal. Jos\'e Cristino 77, Rio de Janeiro, RJ - 20921-400, Brazil}
\author{J.~L.~Marshall}
\affiliation{George P. and Cynthia Woods Mitchell Institute for Fundamental Physics and Astronomy, and Department of Physics and Astronomy, Texas A\&M University, College Station, TX 77843,  USA}
\author{F.~Menanteau}
\affiliation{Department of Astronomy, University of Illinois at Urbana-Champaign, 1002 W. Green Street, Urbana, IL 61801, USA}
\affiliation{National Center for Supercomputing Applications, 1205 West Clark St., Urbana, IL 61801, USA}
\author{R.~Miquel}
\affiliation{Instituci\'o Catalana de Recerca i Estudis Avan\c{c}ats, E-08010 Barcelona, Spain}
\affiliation{Institut de F\'{\i}sica d'Altes Energies (IFAE), The Barcelona Institute of Science and Technology, Campus UAB, 08193 Bellaterra (Barcelona) Spain}
\author{R.~Morgan}
\affiliation{Physics Department, 2320 Chamberlin Hall, University of Wisconsin-Madison, 1150 University Avenue Madison, WI  53706-1390}
\author{A.~Palmese}
\affiliation{Fermi National Accelerator Laboratory, P. O. Box 500, Batavia, IL 60510, USA}
\affiliation{Kavli Institute for Cosmological Physics, University of Chicago, Chicago, IL 60637, USA}
\author{S.~Pandey}
\affiliation{Department of Physics and Astronomy, University of Pennsylvania, Philadelphia, PA 19104, USA}
\author{F.~Paz-Chinch\'{o}n}
\affiliation{Institute of Astronomy, University of Cambridge, Madingley Road, Cambridge CB3 0HA, UK}
\affiliation{National Center for Supercomputing Applications, 1205 West Clark St., Urbana, IL 61801, USA}
\author{A.~A.~Plazas}
\affiliation{Department of Astrophysical Sciences, Princeton University, Peyton Hall, Princeton, NJ 08544, USA}
\author{M.~Rodriguez-Monroy}
\affiliation{Centro de Investigaciones Energ\'eticas, Medioambientales y Tecnol\'ogicas (CIEMAT), Madrid, Spain}
\author{A.~Roodman}
\affiliation{Kavli Institute for Particle Astrophysics \& Cosmology, P. O. Box 2450, Stanford University, Stanford, CA 94305, USA}
\affiliation{SLAC National Accelerator Laboratory, Menlo Park, CA 94025, USA}
\author{S.~Samuroff}
\affiliation{Department of Physics, Carnegie Mellon University, Pittsburgh, Pennsylvania 15312, USA}
\author{E.~Sanchez}
\affiliation{Centro de Investigaciones Energ\'eticas, Medioambientales y Tecnol\'ogicas (CIEMAT), Madrid, Spain}
\author{V.~Scarpine}
\affiliation{Fermi National Accelerator Laboratory, P. O. Box 500, Batavia, IL 60510, USA}
\author{S.~Serrano}
\affiliation{Institut d'Estudis Espacials de Catalunya (IEEC), 08034 Barcelona, Spain}
\affiliation{Institute of Space Sciences (ICE, CSIC),  Campus UAB, Carrer de Can Magrans, s/n,  08193 Barcelona, Spain}
\author{I.~Sevilla-Noarbe}
\affiliation{Centro de Investigaciones Energ\'eticas, Medioambientales y Tecnol\'ogicas (CIEMAT), Madrid, Spain}
\author{M.~Smith}
\affiliation{School of Physics and Astronomy, University of Southampton,  Southampton, SO17 1BJ, UK}
\author{M.~Soares-Santos}
\affiliation{Department of Physics, University of Michigan, Ann Arbor, MI 48109, USA}
\author{E.~Suchyta}
\affiliation{Computer Science and Mathematics Division, Oak Ridge National Laboratory, Oak Ridge, TN 37831}
\author{M.~E.~C.~Swanson}
\affiliation{National Center for Supercomputing Applications, 1205 West Clark St., Urbana, IL 61801, USA}
\author{G.~Tarle}
\affiliation{Department of Physics, University of Michigan, Ann Arbor, MI 48109, USA}
\author{C.~To}
\affiliation{Department of Physics, Stanford University, 382 Via Pueblo Mall, Stanford, CA 94305, USA}
\affiliation{Kavli Institute for Particle Astrophysics \& Cosmology, P. O. Box 2450, Stanford University, Stanford, CA 94305, USA}
\affiliation{SLAC National Accelerator Laboratory, Menlo Park, CA 94025, USA}
\author{T.~N.~Varga}
\affiliation{Max Planck Institute for Extraterrestrial Physics, Giessenbachstrasse, 85748 Garching, Germany}
\affiliation{Universit\"ats-Sternwarte, Fakult\"at f\"ur Physik, Ludwig-Maximilians Universit\"at M\"unchen, Scheinerstr. 1, 81679 M\"unchen, Germany}
\author{J.~Weller}
\affiliation{Max Planck Institute for Extraterrestrial Physics, Giessenbachstrasse, 85748 Garching, Germany}
\affiliation{Universit\"ats-Sternwarte, Fakult\"at f\"ur Physik, Ludwig-Maximilians Universit\"at M\"unchen, Scheinerstr. 1, 81679 M\"unchen, Germany}
\author{R.D.~Wilkinson}
\affiliation{Department of Physics and Astronomy, Pevensey Building, University of Sussex, Brighton, BN1 9QH, UK}

\collaboration{DES Collaboration}
\noaffiliation

\date{\today}

\begin{abstract}

We investigate potential gains in cosmological constraints from the combination of galaxy clustering and galaxy-galaxy lensing by optimizing the lens galaxy sample selection using information from Dark Energy Survey (DES) Year 3 data and assuming the DES Year 1 \metacal sample for the sources. We explore easily reproducible selections based on magnitude cuts in $i$-band as a function of (photometric) redshift, $z_{\rm phot}$, and benchmark the potential gains against those using the well established \redmagic \cite{Redmagic} sample. We focus on the balance between density and photometric redshift accuracy, while marginalizing over a realistic set of cosmological and systematic parameters. Our optimal selection, the \maglim sample, satisfies $i < 4 \, z_{\rm phot} + 18$ and has $\sim 30\%$ wider redshift distributions but $\sim 3.5$ times more galaxies than \redmagic. Assuming a $w$CDM model (i.e. with a free parameter for the dark energy equation of state) and equivalent scale cuts to mitigate nonlinear effects, this leads to $40\%$ increase in the figure of merit for the pair combinations of $\Omega_m$, $w$, and $\sigma_8$, and gains of $16\%$ in $\sigma_8$, $10\%$ in $\Omega_m$, and $12\%$ in $w$.  Similarly, in \LCDM$ $ we find an improvement of $19\%$ and $27\%$ on $\sigma_8$ and $\Omega_m$, respectively.  We also explore flux-limited samples with a flat magnitude cut finding that the optimal selection, $i < 22.2$, has $\sim 7$ times more galaxies and $\sim 20\%$ wider redshift distributions compared to \maglim, but slightly worse constraints.  We show that our results are robust with respect to the assumed galaxy bias and photometric redshift uncertainties with only moderate further gains from increased number of tomographic bins or the inclusion of bin cross-correlations, except in the case of the flux-limited sample, for which these gains are more significant. 
\end{abstract}


\preprint{DES-2019-0482}
\preprint{FERMILAB-PUB-20-566}

\maketitle

\section{Introduction}
\label{sec:intro}

According to the current consensus cosmological model, \LCDM,  dark matter and dark energy make up most of the energy density of the Universe (see e.g. \cite{Frieman2008}).  However, their nature is still unknown and understanding them presents a grand challenge for present-day cosmology. The pillars for the establishment of an accelerating Universe within a \LCDM \, model have been the characterization of  cosmic microwave background fluctuations (CMB) \cite{2003ApJS..148..175S,2018arXiv180706209P} and distance measurements to Type Ia supernovae (SNIa) \cite{1998AJ....116.1009R,1999ApJ...517..565P}. In addition, the study of the large-scale structure (LSS) in our Universe, which carries a wealth of cosmological information, allows us to further constrain these fundamental physics questions (e.g.  \cite{2006PhRvD..74l3507T,2017MNRAS.470.2617A,2019PhRvL.122q1301A,2020JCAP...05..042I,2020JCAP...05..005D} and references therein).

The first cosmology results from ongoing imaging surveys, such as the Dark Energy Survey (DES\footnote{\url{http://www.darkenergysurvey.org/}}) \cite{DESY1_3x2,2019PhRvL.122q1301A}, the Kilo-Degree Survey (KiDS\footnote{\url{http://kids.strw.leidenuniv.nl/}}) \cite{Joudaki2018,vanUitert2018}, and the Hyper Suprime Cam (HSC\footnote{\url{https://www.naoj.org/Projects/HSC/}}) \cite{Hikage2019,Hamana2020}, have demonstrated the feasibility of complex LSS analyses from photometric data and its value and complementarity to the CMB and SNIa in the establishment of a concordance cosmological model.
Consequently, preparations are also under way for the next generation of surveys that will provide high quality imaging data during this decade. The Rubin Observatory Legacy Survey of Space and Time (LSST\footnote{\url{https://www.lsst.org/}}) \cite{LSST}, Euclid\footnote{\url{https://sci.esa.int/web/euclid}} \cite{Euclid}, and the Nancy Grace Roman Space Telescope (Roman\footnote{\url{https://roman.gsfc.nasa.gov/}}) \cite{Spergel2015} complement each other in terms of area, depth, wavelength, and resolution, and will increase the mapped  volume of the Universe by more than one order of magnitude (see e.g. \cite{Eifler2020A,Eifler2020B}). Two of the main cosmological probes from these surveys are galaxy clustering and weak gravitational lensing which we further discuss below.

Weak gravitational lensing refers to the correlated gravitational distortion induced in background galaxy shapes by foreground LSS as their light travels towards us \cite{Bartelmann2001}. This effect is sensitive to the geometry of the Universe and the growth rate of density fluctuations. Hence, information about the cosmological model can be extracted by correlating the observed shapes of galaxies, which is commonly referred to as cosmic shear, or by correlating the positions of galaxies in the foreground (a biased tracer of the LSS) with the shapes of the galaxies in the background, which is often referred to as galaxy-galaxy lensing. The latter can be combined with the auto-correlation of foreground (lens) galaxy positions, a.k.a. galaxy clustering, to break degeneracies with the bias and improve the robustness and constraining power of the cosmological analysis. Such a multi-probe analysis has been carried out by DES in the analysis of its first year of data (DES Y1) \cite{DESY1_3x2}, and by KiDS, combining their shape measurements with spectroscopic foreground (lens) galaxies from the Galaxies And Mass Assembly (GAMA) survey \cite{vanUitert2018} or from the 2-degree Field Lensing Survey (2dFLenS) and the Baryon Oscillation Spectroscopic Survey (BOSS) \cite{Joudaki2018,KiDS_3x2_BOSS_2020}, over the overlapping areas.

When analyzing galaxy clustering (and its combination with galaxy-galaxy lensing) there is a trade-off between selecting the largest galaxy samples to minimize shot noise and selecting samples with the best redshift accuracy, which generally include only a small subset of galaxies. In this paper we investigate the potential gains in cosmological constraints that can be obtained by optimizing the selection of the lens galaxy sample in a combined galaxy clustering and galaxy-galaxy lensing analysis (hereafter $2\times2$pt). We choose to not include cosmic shear in this work given that the only impact would be an overall increase of the constraining power for all cases, independently of the lens sample considered. Note that, as a consequence, the relative improvements in cosmological constraints in a $3\times2$pt analysis (i.e., when including shear) will be likely smaller than the results presented here.

In order to define samples with accurate redshift estimates from photometric data, a common choice is to use luminous red galaxies (LRGs), which are characterized by a sharp break at 4000$\AA$ \cite{Eisenstein2001,2007MNRAS.378..852P} and a remarkably uniform spectral energy distribution. They also correlate well with clusters. These features allow the selection of this sample of galaxies from the general population, as well as the estimation of their redshifts with high accuracy. The  approach taken in the DES Y1 analysis \cite{DESY1_3x2} consisted of selecting the lens galaxies in terms of optimal photometric redshift (photo-$z$) accuracy\footnote{Note that, in practice, this also translates into a robust and simple characterization of redshift distributions, which otherwise is a difficult task.} using the \redmagic algorithm \cite{Redmagic} which relies on the calibration of the red-sequence in optical clusters. A similar selection of red-sequence galaxies has been carried out recently by KiDS, combining their broad-band optical catalog with near-infrared photometry from the VISTA Kilo-degree Infrared Galaxy (VIKING) survey \cite{KiDS_LRGs_2020}. Selections of LRG's in photometric data, based on color and magnitude cuts, have been done also for measurements of Baryon Acoustic Oscillations \cite{2007MNRAS.378..852P, 2019MNRAS.482.2807C,2020arXiv200513126S}.

An alternative choice is to select all galaxies up to a limiting magnitude. This can lead to galaxy samples that reach  higher redshifts with a much higher number density, at the expense of lower photo-$z$ accuracy. Flux-limited samples have been used, for example, in the DES Science Verification analysis \cite{Crocce2016} and, previously,  in the galaxy clustering measurements from Canada-France-Hawaii Telescope Legacy Survey (CFHTLS) data \cite{Coupon2012}. Both analyses were very similar in  terms of depth, photometry, and area, and the samples were defined with the same cut in apparent magnitude: $i<22.5$. More recently, a flux-limited sample has been considered in the galaxy clustering measurements from HSC data \cite{Nicola2020}, in which the authors select galaxies with a limiting magnitude $i<24.5$ and study their properties such as large-scale bias.  This kind of galaxy selection is simple and easily reproducible in different data-sets and, as a consequence, the properties of the sample can be well understood. For instance, in \cite{Crocce2016} the authors show that the redshift evolution of the linear galaxy bias of their sample matches the one from CHFTLS \cite{Coupon2012}, and  this redshift evolution  also agrees well with that from HSC data \cite{Nicola2020}. However, to our knowledge this type of selection has not yet been used to derive cosmological constraints.

In this work, we follow this approach and consider flux-limited samples as an alternative to the LRG \redmagic sample selected from the third year of DES data (DES Y3), aiming to optimize the lens galaxy selection to extract the maximum amount of cosmological information. We will then consider this optimal sample as one of the lenses of the upcoming DES Y3 analysis,  not only because of potential improved constraints but also as a test of the robustness of the cosmological results given the characteristics of the lens galaxy sample such as its redshift extent, bias, photo-$z$ characterization or density. In the follow-up paper \citep[][in prep.]{DESY3maglim}, we will obtain cosmological constraints from the galaxy clustering and galaxy-galaxy lensing measurements of this sample. We will also validate the redshift distributions, the treatment of photometric uncertainties, the scale cuts, and the modeling pipeline.

This paper is organized as follows. In Sec.~\ref{sec:DESY3data} we describe the DES data used and the sample selections we consider throughout. In Sec.~\ref{sec:forecasting-methodology} we detail our methodology to infer cosmological parameter constraints including the theory modeling, the parameter space (cosmological and systematic), and the scale cuts. In Sec.~\ref{sec:sampleoptimization} we discuss the optimization process, which reflects the core of our results. In Sec.~\ref{sec:sample-comparison} we describe the optimal samples and compare their properties and cosmological constraints obtained from  Monte-Carlo Markov chains (MCMC\footnote{Technically, in this work we use a Monte-Carlo (MC) method instead of other traditional MCMC techniques. However, since the end product of these two kinds of methods is equivalent, we employ the `\emph{MCMC}' acronym because it is a more established term in the literature.}) to provide realistic Y3 simulated analysis. In Sec.~\ref{sec:analysis-choices} we study the performance of the optimized samples for different analysis choices such as the binning strategy, assumed galaxy bias or photo-$z$ error priors. We finish in Sec.~\ref{sec:conclusions} presenting our conclusions.

\section{DES Y3 Data}
\label{sec:DESY3data}


DES \cite{Flaugher2015} is an imaging survey of $\sim 5000 \,\sqdeg$ of the southern sky, using a 570 megapixel camera \citep[DECam;][]{Flaugher2015} mounted on the 4 m Blanco telescope at the Cerro Tololo Inter-American Observatory (CTIO)  in Chile in five broadband filters, $grizY$.  The main goal of DES is to determine the dark energy equation of state parameter $w$ and other key cosmological parameters.  In this work we use data  from the first three years of observations (Y3), which were taken from August 2013 to February 2016.

The  catalog that will be used for the cosmological analysis of Y3 data, the \gold catalog,  is described more extensively in \cite{DESy3gold} and it is based on the coadded catalog from the first three years of data, which was released publicly as the DES Data Release 1 (DR1)\footnote{Available at \href{https://des.ncsa.illinois.edu/releases/dr1}{https://des.ncsa.illinois.edu/releases/dr1} }. The DES DR1 is the first DES catalog that spans the whole footprint,  
and it is described in \cite{DESDR1}, alongside with the details of the Data Management pipeline in \cite{Morganson2018} and photometric calibrations in \cite{Burke2018}. 
The source catalog was built using \sextractor \citep{Bertin:1996} detecting objects on the $grizY$ co-added images up to a 10-$\sigma$ limiting magnitude of 
$g=24.3$, $r=24.0$, $i=23.3$, $z=22.6$ and $Y=21.4$ mag. (see Table 2 in \cite{DESy3gold}).
In this work, however, we only use the DES \gold catalog for the lens samples; for the sources, we employ the \metacal source sample \cite{Troxel2018} built from the DES \code{Y1 GOLD} catalog \cite{y1gold}. 

The photometry in Y3 is derived through the Multi-Object Fitting pipeline \cite{y1gold} and its variant  Single-Object Fitting (\sof),  which eliminates the multi-object light subtraction speeding up the process with negligible impact on performance. In this paper we use \sof magnitudes for sample selection and as input to the photometric redshift codes. In particular, we select the samples from the \texttt{Y3\_GOLD\_2\_2} catalog using the SOF magnitudes corrected for Galactic extinction and other minor adjustments (\code{SOF\_CM\_MAG\_CORRECTED}) and we remove stellar contamination from our samples by using the default star-galaxy separation method from  \cite{DESy3gold}  (\code{EXTENDED\_CLASS\_MASH\_SOF}$=3$), which reduces the stellar contamination to less than 2\%.
The \gold catalog contains $\sim 388$ million objects detected in co-added images covering $\sim 5000$ $\sqdeg$ in the DES $grizY$ filters.


As part of the \gold dataset, three standard photometric redshift codes were run (one template fitting, \bpz \cite{Benitez:2000} and two machine learning, \texttt{ANNz2} \citep{annz2} and \dnf \cite{DNF2016}).
In this paper we rely exclusively (aside from \redmagic) on the \dnf run based on \sof photometry that is provided as part of the \gold catalog. The Directional Neighborhood Fitting (\dnf) algorithm creates an approximation of the redshift of the object through a nearest-neighbors fit in a hyperplane in color and magnitude space using a reference training set from a spectroscopic database.  The database of spectra is described in \cite{Gschwend:2018} and includes $\sim 220$ thousand spectra matched to DES objects from 24 different spectroscopic catalogs, such as SDSS DR14 \citep{sdssdr14}, the OzDES program \citep{ozdes} and VIPERS \citep{VIPERS:2014}, among others. In the case of \dnf about half of these spectra are used for training and the rest for performance validation. The performance of the different photometric redshift runs is discussed in \cite{DESy3gold}, where it is found that \dnf outperforms the other methods in standard metrics such as width and biases of photometric redshift error distributions.  In addition, \dnf also provides the redshift of the actual nearest-neighbor within the reference training sample, which together with the approximated redshift estimate $z_{\rm phot}$ serves as an internal metric for the \photoz redshift error per object.

\subsection{Sample selections}
\label{sec:sample-selections}

As noted in the introduction, we use different kinds of lens samples defined from DES Y3 data. Aside from a \redmagic sample, we define two types of flux-limited samples. The first one consists of an overall apparent magnitude limit, similar to what has been commonly used in previous analyses, and the second one ($\maglim$) is a sample defined with a magnitude cut varying linearly with redshift.  This avoids selecting red objects through explicit color cuts since that would  mimic \redmagic. Thus, given the DNF photo-$z$ values for \maglim, both of these definitions lead to selections that are easy to implement and reproduce in practice. 
Our samples are hence defined mainly in terms of their luminosity (as a function of redshift).  In the following, we describe our sample selection criteria, their photometric redshift estimates, and the effective survey area and angular mask applied to them. Both flux-limited samples are optimized in Sec. \ref{sec:sampleoptimization}.

\subsubsection{Flux-limited sample}
\label{sec:flux-limited-sample}

Flux-limited samples are defined with a flat apparent magnitude cut on the $i$-band, $i<a$ with $a$ being some constant, because generally it is the magnitude with the best signal-to-noise ratio per object over the redshift range considered. This type of sample has been used in various analyses in the past, e.~g. the galaxy clustering analysis of DES Science Verification data \cite{Crocce2016}, and also in CFHTLS \cite{Coupon2012} and HSC \cite{Nicola2020}. In particular, \cite{Tanoglidis2019} considers this approach, using DES Y1 data, to study the trade-off between number density and \photoz accuracy and its impact in terms of cosmological constraints from galaxy clustering with fixed bias parameters. Therefore, it is interesting to consider this type of sample here, and compare it with the other two samples, $\maglim$ and \redmagic, described next. 

\subsubsection{\maglim \,sample}
\label{sec:maglim-sample}

One possible disadvantage of selecting all galaxies up to a fixed limiting magnitude is that at low redshift the selection includes a higher number of less luminous (mostly blue) galaxies, degrading the \photoz accuracy as a result. For this reason, here we explore a different galaxy selection that serves as an intermediate scenario in terms of number density and photometric redshift accuracy. In particular, we consider samples selected with a limiting magnitude that varies across redshift, of the type $i<a z_{\rm phot} +b$, with $a$ and $b$ arbitrary numbers and $z_{\rm phot}$ being the \dnf \photoz estimate. Effectively this selects brighter galaxies at low redshift while including fainter galaxies as redshift increases.  Additionally,  we remove the brightest objects (including stellar contamination from binary stars) by setting $i > 17.5$. 

\subsubsection{\redmagic}

This galaxy sample, which will be described more extensively in \citep[][in prep.]{y3-galaxyclustering},  is generated by the \redmagic algorithm \cite{Redmagic} run on DES \gold data. The \redmagic algorithm selects LRGs in such a way that photometric redshift uncertainties are minimized. This algorithm fits every galaxy to a red-sequence template, and only includes in the selection galaxies that are bright enough (above a certain luminosity threshold $L_{\min}$), and that have a good enough fit to the red-sequence template using the assigned photometric redshift ($\chi^2\le \chi^2_{\max}$).
In addition, it is required that  the resulting sample has constant comoving density as a function of redshift. The red-sequence template is generated by the training of the $\redmapper$ cluster finder \cite{Rykoff2014,Rykoff2016}. Reference luminosities are defined as a function of $L_{\ast}$, computed using a Bruzual and Charlot \cite{Bruzual2003} model for a single star-formation burst at $z=3$, as described in \cite{Rykoff2016}. Naturally, increasing the luminosity threshold provides a higher redshift sample as well as decreasing the comoving number density.

Two \redmagic samples are generated from the Y3 data, equivalent to the ones from Y1 \cite{ElvinPoole2018}, and referred to as high-density and high-luminosity. The corresponding luminosity thresholds and comoving densities are, $L_{\min}=0.5L_{\ast}$, and $1.0L_{\ast}$, and $\bar{n}=10^{-3},$  and $4\times 10^{-4}$ galaxies/($h^{-1} \Mpc$)$^3$, where $h$ is the reduced Hubble constant. The combined \redmagic sample we use in this work consists of high-density galaxies at redshifts $z<0.65$, and high-luminosity galaxies in the range $0.65 < z < 0.95$. The \redmagic algorithm produces  best-fit redshifts, which we use as the estimated photometric redshifts. These photometric redshifts are particularly accurate, with an uncertainty $\sigma_z/(1+z) < 0.02$, see  Figure~\ref{fig:sigz} for the dependency of this uncertainty with redshift.

\subsection{Sample comparison}

\begin{figure}
	\begin{center}
		\includegraphics[width=\linewidth]{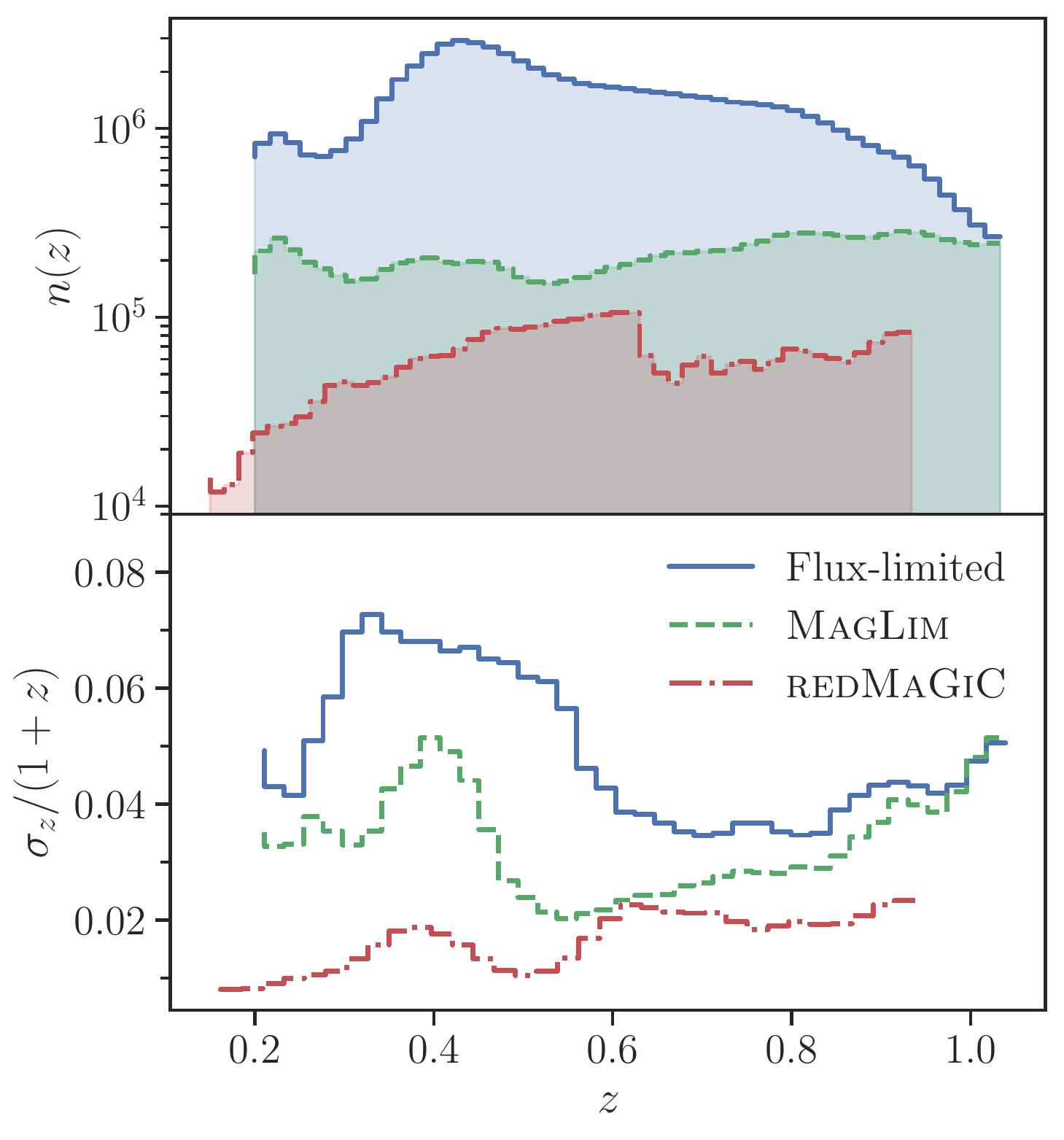}
		\caption{Galaxy counts (top panel) and mean \photoz error $\sigma_z/(1+z)$ (bottom panel) as a function of photometric redshift for three  cases of the lens samples considered in this work (see text for details). }
		\label{fig:sigz}
	\end{center}
\end{figure}

In Figure~\ref{fig:sigz} we show the galaxy counts (top panel) and the mean photo-$z$ error (bottom panel) as a function of the photometric redshift for  the three types of samples we discussed above. For the flux-limited sample we show $i < 22.2$ while for \maglim $i < 18+4 z_{\rm phot} $, where $z_{\rm phot}$ is the \dnf photometric redshift  estimate.  The mean \photoz error $\sigma_z$ is obtained in different ways depending on the galaxy sample. In the case of the $\redmagic$ sample, $\sigma_z$ corresponds to the redshift uncertainty provided by the \redmagic algorithm. For $\maglim $ and flux-limited samples, however, $\sigma_z/(1+z)$ is the 68\% confidence interval of values in the distribution of $(z_{\rm phot} - z_{\rm true})/(1+z_{\rm true})$ around its median value, where  $z_{\rm true}$ corresponds to the \dnf nearest-neighbor redshift. Figure ~\ref{fig:sigz} shows that while the flux-limited sample has many more galaxies (especially at low redshift), the photometric redshift accuracy is far from optimal, with $0.04 < \sigma_z/(1+z)< 0.07$. With the $\maglim$ sample we exclude from the selection the faintest / bluest galaxies that have worst \photoz, while still managing to get a sample with several times the number density of \redmagic. The \photoz accuracy, thus, improves with respect to the flux-limited sample, with $0.02 < \sigma_z/(1+z) < 0.05$.  Note also that the maximum redshift range (before the sample starts being incomplete and the \photoz error degrades) is $z_{\rm max}\sim 1.05$ for \maglim compared to $z_{\rm max}  \sim 0.95$ for \redmagic.

\subsection{Tomographic binning and redshift distributions}
\label{sec:data-photo-z}

In the rest of the paper we will derive cosmological constraints after dividing the samples in tomographic bins and using estimates for the distribution of true redshifts per bin. 

The estimate for galaxy redshifts (\photoz) used for tomographic binning and galaxy selection for the $\maglim$ and flux-limited samples is derived using  the predicted value in the fitted hyper-plane from the \dnf code. In turn, it has been shown that the stacking of the nearest-neighbor redshift allows the method to replicate science sample redshift distributions accurately \cite{2018MNRAS.478..592H,2018MNRAS.477.1664G}, and results and performance in \gold are similar to those found previously by \cite{lima}. In follow-up papers we will investigate the performance of this approach for \maglim against direct calibration with spectroscopic fields \cite{DESY3maglim} and clustering redshifts \cite{Cawthon2020} in more detail. Hence for the estimates of the redshift distribution of galaxies in each tomographic bin, $n(z)$, we use the stacking of the nearest-neighbor redshifts of the galaxies in the sample.

For the \redmagic sample we assume that the redshift probability distribution function (PDF) for each galaxy is a Gaussian distribution with mean given by the \redmagic best-fit redshift and standard deviation $\sigma_z$. We then obtain an overall estimate of the redshift distributions by stacking these Gaussian PDFs \cite{Redmagic,ElvinPoole2018}.

\subsection{Survey area and angular mask}
\label{sec:survey-area}

The footprint of the DES \gold catalog amounts to 4946 $\sqdeg$. For cosmology analyses, additional masking is applied to remove bright stars and other foreground objects, and also regions of the footprint that have some deficiency in the source extraction of photometric measurement (a.k.a. \emph{bad regions}). As a result, the effective area is reduced by 659.68~$ \sqdeg$ \cite{DESy3gold}.  

\begin{figure}
\begin{center}
\includegraphics[width=\linewidth]{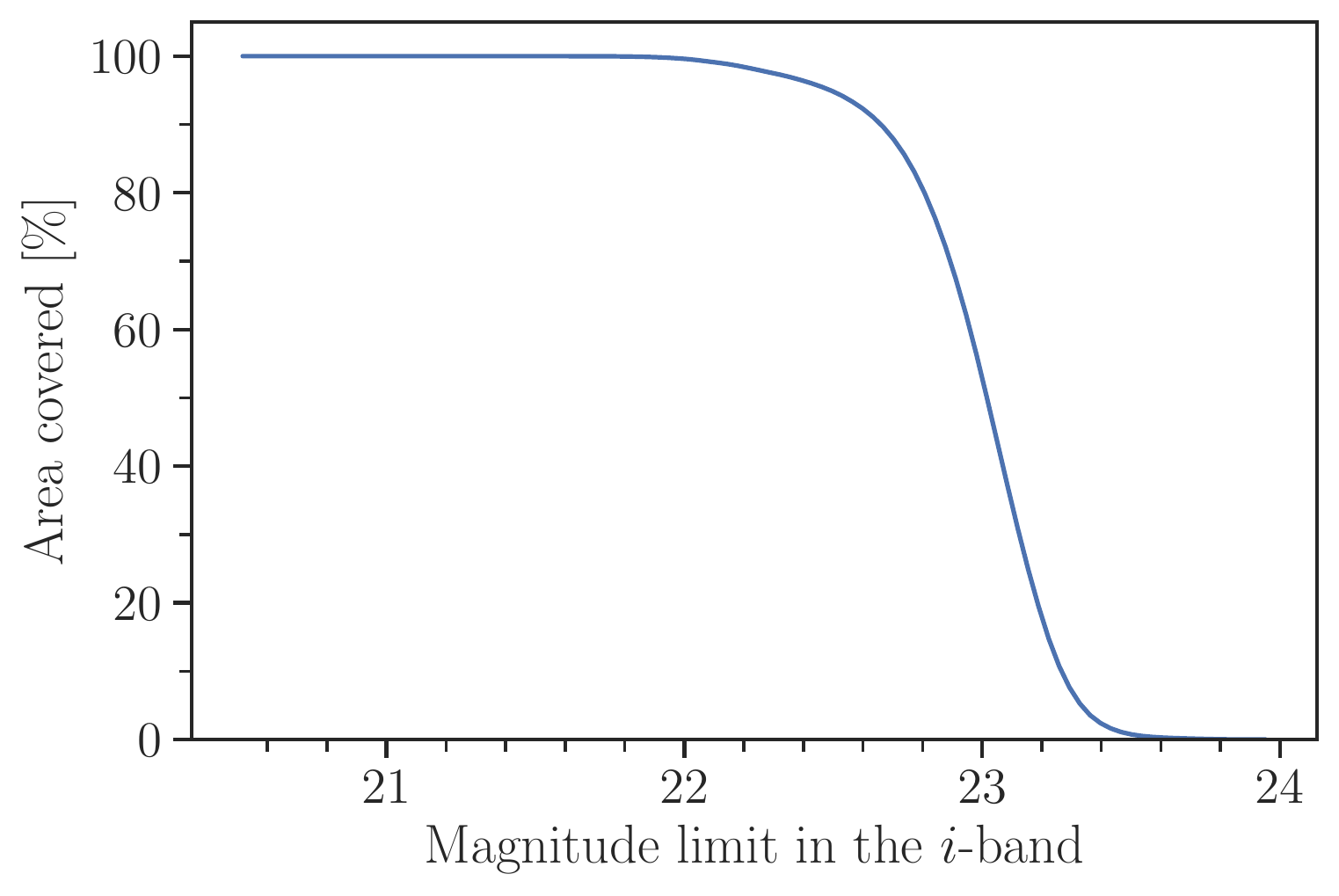}
\caption{Percentage of survey area as a function of the limiting magnitude in the $i$-band. The reference full area includes all the baseline quality cuts corresponding to the  DES  \gold dataset.}
\label{fig:area}
\end{center}
\end{figure}

Then, for a given galaxy sample, we mask the regions that are too shallow in order to have a homogeneous depth across the footprint. 
In Figure~\ref{fig:area} we show the fractional survey area as a function of the limiting magnitude reached in that area in the $i$-band. Samples with an overall limiting magnitude of $i = 22$ or lower will be complete over $100\%$ of the footprint. If we increase the limiting magnitude to incorporate more objects into the sample, then the regions of the sky that are too shallow would need to be masked in order to achieve a homogeneous depth. Therefore, there is a trade-off between imposing limits at higher magnitudes and preserving the survey area. In Sec.~\ref{sec:sampleoptimization} we vary a range of limiting magnitudes in order to optimize the samples and decide not to consider those selections with $i >22.75$, at which point we would need to mask $\sim10\%$ or more of the sky area.

The samples that we find to be optimal in terms of 2$\times$2pt cosmological constraints are complete in regions of the survey deeper than $i = 22.2$ magnitudes. Therefore, we will consider such regions as our baseline footprint.  This implies masking out about $\sim1\%$ of the area. A similar masking is applied for the \redmagic sample. We use depth information from the \redmagic catalogs to mask out the regions in the footprint that are too shallow.
Since we want to compare the cosmological constraints obtained from the optimal samples with the \redmagic sample, we then combine these two masks resulting in a unique mask that is applied to both. Using the same mask for both samples reduces the area by an additional $\sim 100$ $\sqdeg$,  yielding a final effective area of $4182$ $\sqdeg$.
For simplicity, we use the same mask for all sample selections. We note that this is optimistic for those samples in Sec.~\ref{sec:sampleoptimization} with limiting magnitudes larger than 22.2.

\section{Forecasting Methodology}
\label{sec:forecasting-methodology}

In what follows, we describe the methodology employed for sample optimization. For each magnitude cut considered we access the catalog and apply the sample selection, which leads to a given number density and redshift distribution per tomographic bin. From these, we produce theory data-vectors and covariances that are subsequently used to derive cosmological parameter constraints following the forecasting methodology that we present next.

\subsection{Likelihood exploration}
\label{sec:likelihood-exploration}

In order to investigate the potential gains in cosmological constraints,  we run simulated likelihood analyses with Fisher matrix \cite{Tegmark1997,Tegmark1997B} and MCMC methods. 
The Fisher matrix is commonly used for forecasting constraints because it is fast to compute and provides an approximation for the covariance matrix of the parameters. However, since the Fisher matrix is a local approximation of the likelihood, it can provide inaccurate results for non-Gaussian posterior distributions, as is the case when there are degeneracies between parameters (see e.g. \cite{Wolz2012}).
A more robust approach for forecasting is possible by sampling the full posterior distributions using an MCMC approach.  

We sample the posterior in the $n$-dimensional parameter space by computing the likelihood at every step, where $n$ is the number of parameters ($\vec{p}$) we vary in our analysis (see Table~\ref{tab: priors}). We assume the likelihood to be Gaussian,
\begin{equation}
\begin{gathered}[l]
\hspace*{-2cm} \ln \mathcal{L}(\vec{d}|\vec{m}(\vec{p})) = \\
 -\frac{1}{2} \sum^N_{ij} \left(d_i -m_i(\vec{p})\right) C^{-1}_{ij} \left(d_j - m_j(\vec{p}) \right).  
\end{gathered}
\end{equation}
Here $N$ is the number of data points,  $\vec{m}(\vec{p})$ are the theoretical predictions as a function of the parameters we allow to vary, $\vec{d}$ is the noiseless theory data vector (the set of  theoretical predictions evaluated at the fiducial cosmology), and $C$ is the covariance matrix, also evaluated at the fiducial cosmology (see Table~\ref{tab: priors}).
The posterior distribution of the  parameters is given by: 
\begin{equation}
P(\vec{m}(\vec{p})|\vec{d}) \propto \mathcal{L}(\vec{d} | \vec{m}(\vec{p})) P_{\rm prior} (\vec{p}),
\end{equation}
where $P_{\rm prior}(\vec{p})$ is the prior on the parameters. The Fisher matrix is defined as the expectation value of the curvature of the log-likelihood evaluated at the maximum likelihood point, i.e. the fiducial values of the parameters $\vec{p}_0$
\begin{equation}
F_{ij}\equiv - \left\langle \left. \dfrac{\partial^2 \log  \mathcal{L} }{\partial p_i \partial p_j} \right\vert_{\vec{p}=\vec{p}_0} \right\rangle.
\label{eq:fisher-matrix}
\end{equation}
 We can include Gaussian priors by adding a prior matrix 
	\begin{equation}
	F_{ij}^P = \delta_{ij}\dfrac{1}{(\sigma_i^P)^2},
	\label{eq: fisher-priors}
	\end{equation}
where $\sigma_i^P$ is the standard deviation on the parameter $p_i$ assumed as a prior.
According to the Cramér-Rao inequality, the Fisher matrix gives a lower bound on the error $\sigma$ on a parameter $p_i$, 
\begin{equation}
\sigma(p _i) \geq \sqrt{(F^{-1})_{ii}}.
\end{equation}
A commonly used metric to measure the constraining power of a given data set is the figure of merit (FoM). The FoM for a subset of cosmological parameters $p$ is defined as
\begin{equation}
\mathrm{FoM}_p = \dfrac{1}{\sqrt{\det \left[ (F^{-1})_{p} \right]}},
\end{equation}
where $(F^{-1})_{p} $ is the selection on $(F^{-1})$ of the rows and columns corresponding to the subset of parameters $p$.  An intuitive way to understand the FoM is to consider a subset of two parameters. In that case, the FoM is inversely proportional to the area of the confidence ellipse of these two parameters.

One of the most important factors for the reliability of a Fisher matrix is the stability of the numerical derivatives (see e.g. \cite{Wolz2012,Camera2017,Euclid_Fisher_forecasts}). The computation of the derivatives involves evaluating the likelihood at several points in the vicinity of the fiducial values of the parameters, assuming a given step size. The problem is that if the step size is too large, the numerical derivative may not be accurate. On the other hand, if the step size is too small, the derivative estimate will be unreliable due to numerical instabilities. For this purpose, when computing a Fisher matrix, we iteratively vary the step size for each parameter until we reach a certain tolerance. In the following, we explain the details of this process. We first compute the derivatives at an initial step size of 0.01 (1\%) in units of the range of each parameter. Assuming a maximum step size $s_{\max}=0.05$ (which is a reasonable boundary according to \cite{Fishervalidation2020}), in each iteration we vary the step sizes to the minimum value between $s_{\max}$ and the predicted $\sigma$ error on that parameter: $s_{\mathrm{new}}=\min(s_{\max},\, \sigma(p_i))$. The algorithm converges when the differences in the sigma errors $\sigma(p_i)$ are below a tolerance of 0.01 and the differences in the predicted covariance matrix of the parameters are below $10^{-4}$.  

Another important factor is the treatment of priors in the Fisher matrix estimation. In general, when analyzing data we assume wide flat priors for the cosmological parameters in order to avoid having cosmological results that depend on the priors assumed. However, as mentioned before, the Fisher matrix will fail to estimate the posterior distributions in the presence of non-Gaussianities, which can lead to confidence contours that extend beyond the physically meaningful parameter range. In order to address this, we apply wide Gaussian priors for the parameters listed with flat priors in Table~\ref{tab: priors}, assuming in Eq.~\eqref{eq: fisher-priors} a standard deviation equal to half the limits $[a, b]$ of the parameter range in Table~\ref{tab: priors}: $\sigma_i^P = (b-a)/2$. This was the approach taken in \cite{Camera2017}, which resulted in a good agreement between Fisher and MCMC.  In the case of nuisance parameters with Gaussian priors, we just assume as $\sigma_i^P$ the $\sigma$ values listed in Table~\ref{tab: priors}.

Even though we have taken measures to ensure our Fisher matrices are reliable, the predicted constraints $\sigma(p_i)$  will still have an uncertainty of order $10\%$ with respect to other Fisher codes and MCMC methods  \cite{Euclid_Fisher_forecasts,Camera2017,Fishervalidation2020}. 
For this reason, in Sec.~\ref{subsec:cosmological-constraints}    we compare a representative set of our Fisher forecasts against the constraints coming from a full MCMC sampling of the posterior. 
Nevertheless, we rely on the Fisher matrix for most of our forecasts, with the exception of  Sec.~\ref{subsec:cosmological-constraints}, in which we show the MCMC constraints for $\redmagic$ and the optimal samples.

In this paper we use  \cosmosis \footnote{{\href {}https://bitbucket.org/joezuntz/cosmosis/}} \cite{Zuntz2015}  to compute the Fisher matrices. For the MCMC simulated likelihood analyses,  we sample the posterior distribution using the  \textsc{Multinest} \cite{Feroz2009} wrapper in \cosmosis.

\subsection{Theory Modelling}
\label{sec:theory}

In this section, we describe the model we use to characterize galaxy clustering and galaxy-galaxy lensing and their covariance matrix. As seen in Sec.~\ref{sec:likelihood-exploration}, we use these to extract cosmological information from a given data vector that, in our case, is a noiseless theoretical prediction at the fiducial cosmology. The model depends upon both cosmological parameters and astrophysical and systematic nuisance parameters (see Sec.~\ref{sec:params-and-priors}). In Appendix~\ref{sec:cov-comp}, we validate the numerical implementation of our covariances by comparing the constraints coming from two different covariance codes.

\subsubsection{Observables}

The observables we consider in the simulated likelihood analyses are the galaxy clustering and galaxy-galaxy lensing two-point angular correlation functions, i.e. the correlations in the positions of the lens galaxies, and the correlation between these positions and the source galaxy shears.

Under the Limber approximation \cite{Loverde2008}, we can construct their respective angular power spectra as a function of multipole $\ell$ in the following way,

\begin{equation}
\begin{gathered}[l]
C^{ij}_{\delta_g \delta_g} (\ell) = \int d\chi \frac{q_{\delg}^i \left(\frac{\ell + \frac{1}{2}}{\chi},\chi \right) \, q_{\delg}^j \left(\frac{\ell + \frac{1}{2}}{\chi},\chi \right) }{\chi^2} \, \\
 \times \, P_{\rm NL} \left(\frac{\ell+\frac{1}{2}}{\chi}, z(\chi)\right) ,
 \end{gathered}
 \label{eq:c_ell_galaxy}
 \end{equation}
 \begin{equation}
 \begin{gathered}[l]
C^{ij}_{\delta_g \kappa} (\ell) = \int d\chi \frac{q_{\delg}^i \left(\frac{\ell + \frac{1}{2}}{\chi},\chi \right) \, q_{\kappa}^j (\chi) }{\chi^2} \, \\
 \times \, P_{\rm NL} \left(\frac{\ell+\frac{1}{2}}{\chi}, z(\chi)\right),
 \end{gathered}
 \label{eq:C_ell_gglensing}
\end{equation}
where $P_{\rm NL}(k, z)$ is the non-linear matter power spectrum,  and $q_{\delg}^i$ and $q^j_{\kappa}$ are respectively the density kernel in the redshift bin $i$ from the lens sample, and the lensing efficiency in the redshift bin $j$ from the source sample. These kernels depend, respectively, on the redshift distributions of lens ($n^i_{\delta_g}(z)$) and source ($n^i_{\kappa}(z)$) galaxy samples normalized by their respective total number densities in that redshift bin  ($\bar{n}^i_{\delta_g} $ for the lenses and $\bar{n}^i_{\kappa} $ for the sources), and can be expressed as a function of the comoving distance $\chi$ in the following way,

\begin{equation}
q_{\delg}^i (k,\, \chi) = b^{i}(k,z(\chi)) \, \frac{n_{ \delta_g}^i(z(\chi))}{\bar{n}_{\rm g}^i} \, \frac{dz}{d\chi},
\label{eq:density kernel}
\end{equation}
\begin{equation}
\begin{gathered}[l]
q_{\kappa}^i (\chi) = \frac{3H_0^2 \Omega_{\rm m}}{2c^2} \, \frac{\chi}{a(\chi)} \, \\
\times \, \int_{\chi} ^{\chi_{\rm h}} \der \chi' \frac{n_{\kappa}^i(z(\chi')) \der z / \der \chi'}{\bar{n}_{\kappa}^i} \frac{\chi' - \chi}{\chi'} \, , 
\end{gathered}
\end{equation}
where $H_0$ is the Hubble constant, $c$ is the speed of light, $a$ is the scale factor, and $b^{i}(k,z)$ is the galaxy bias, a nuisance parameter that we vary in our analysis (see Sec.~\ref{sec:params-and-priors}). We adopt a linear galaxy bias model (independent of the scale $k$), with a single galaxy bias $b_i$ parameter for each redshift bin.
	
Under the flat-sky approximation, the galaxy clustering and galaxy-galaxy lensing angular two-point correlation functions can be computed from the angular power spectra from Eqs.~\eqref{eq:c_ell_galaxy} and \eqref{eq:C_ell_gglensing} in the following way, 

\begin{equation}
	w^{ij}(\theta) = \int \dfrac{\der \ell\,  \ell}{2\pi} J_0(\ell\theta) C^{ij}_{\delta_g \delta_g} (\ell),
\end{equation}
\begin{equation}
\begin{gathered}[l]
\gamma_{\rm t}^{ij} (\theta) = (1+m^j) \int \dfrac{\der \ell\,  \ell}{2\pi} J_2(\ell\theta) C^{ij}_{\delta_g \kappa} (\ell),
\end{gathered}
\end{equation}
 where $J_n$ is  the $n$-th order Bessel function of the first kind, and $m^j$ is the multiplicative shear bias, a nuisance parameter introduced to take into consideration potential biases in the inferred shear.

In most of this work we restrict $w(\theta) $ to auto-correlations within each redshift bin, i.e. we just consider $w^{ii}$. However, in Sec.~\ref{sec:tomographic-binning} we test the impact of including galaxy clustering cross-correlations between redshift bins in our analysis.

In addition to the galaxy shear induced by gravitational lensing, galaxy shapes can also be intrinsically aligned as a result of their formation and evolution in the same large-scale structure environment.  The impact of intrinsic alignments (IA) can be modeled using a power spectrum shape and an amplitude $A(z)$. We assume the non-linear alignment model (NLA) \cite{Hirata2004,Bridle2007} for the IA power spectrum, which impacts the lensing efficiency in the following way,

\begin{equation}
q_{\kappa}^i (\chi) \rightarrow q_{\kappa}^i (\chi) - A(z(\chi)) \dfrac{n_{\kappa}^i(z(\chi)) }{\bar n_{\kappa}^i} \dfrac{\der z}{\der \chi}.
\end{equation}

We model the IA amplitude assuming a power-law scaling with redshift, 

\begin{equation}
A(z) = A_{\mathrm{IA},0}\left(\dfrac{1+z}{1+z_0}\right)^{\alpha_{\mathrm{IA}}} \dfrac{C_1 \rho_{\textrm{crit}}}{D(z)},
\end{equation}
where $D(z)$ is the linear growth factor. The pivot redshift is chosen to be approximately the mean redshift of the sources, $z_0 = 0.62$, and $C_1 \rho_{\textrm{crit}} = 0.0134$ is a normalization derived from \textsc{SuperCOSMOS} observations \cite{Bridle2007}. Therefore, the IA model assumed adds two extra nuisance parameters in our analysis: $A_{\mathrm{IA},0}$ and $\alpha_{\mathrm{IA}}$.

We note that magnification, which we do not include in our modeling, will be significant when using  flux-limited samples on a real data analysis. Ref.~\citep[][in prep.]{y3-2x2ptmagnification} will show the measurement and validation of the magnification coefficients for both \redmagic and the optimal sample resulting from this work. These coefficients will be included in the DES Y3 analysis to avoid biases on the cosmological constraints. However, the constraining power is only slightly degraded when marginalizing over the magnification coefficients \cite{y3-2x2ptmagnification}. Therefore, our conclusions are not affected by the neglect of magnification effects. 

We calculate the power spectrum using the Boltzmann code CAMB\footnote{See \texttt{camb.info}.} \cite{Lewis2000,Howlett2012} with the \textsc{Halofit} extension to nonlinear scales \cite{Smith2003,Takahashi2012} and the neutrino extension from \cite{Bird2012}. We use $\cosmosis $ to compute the galaxy clustering and tangential shear two-point functions.

\subsubsection{Covariance}

Following the notation in Refs.~\cite{Krause2017,Fang2020}, in the flat sky limit, the real space covariance of two angular two-point functions $\Xi,\, \Theta \in \{w, \gamma_t\}$ at angles $\theta$ and $\theta'$ is related to the covariance of the angular power spectra by

\begin{equation}
\begin{gathered}[l]
\mathrm{Cov}(\Xi^{ij}(\theta),\, \Theta^{km}(\theta')) = \\
 \frac{1}{4\pi^2} \int \der \ell \, \ell J_{n(\Xi)} (\ell \theta) \int \der \ell' J_{n(\Theta)} (\ell' \theta') \\
\times \left[ \mathrm{Cov}^G(C_{\Xi}^{ij}(\ell), C_{\Theta}^{km}(\ell')) + \mathrm{Cov}^{NG}(C_{\Xi}^{ij}(\ell), C_{\Theta}^{km}(\ell')) \right] ,
\end{gathered}
\end{equation}

with $C_{\gamma_t} \equiv C_{\delta_g\kappa}$ from Eq.~\eqref{eq:C_ell_gglensing}, and $C_w \equiv C_{\delta_g \delta_g}$ from Eq.~\eqref{eq:c_ell_galaxy}, and where the order of the Bessel function is $n=0$ for $w$, and $n=2$ for $\gamma_t$. The indices $i, j, k, m$ denote the redshift bins.
All two-point functions are evaluated in 20 log-spaced angular bins over the range $2.5' < \theta < 250'$. This yields a 500$\times$500 covariance matrix if the lens sample is split in 5 tomographic bins (which is the fiducial case for the flux-limited and $\redmagic$ samples), and the size increases by 100 for each additional tomographic bin.
The non-Gaussian covariance $\mathrm{Cov^{NG}}$ consists of a connected four-point correlation contribution \cite{Cooray2002,Takada2009} and a super-sample contribution \cite{Takada2013}. In the Gaussian covariance $\mathrm{Cov}^G$ \cite{Hu2004} different harmonic modes $\ell$ are uncorrelated, so its harmonic transform reduces to a single integral. The Gaussian covariance has terms related to cosmic variance,  shot noise ($\propto 1/\bar{n}^i$, with $\bar{n}^i$ being the mean number density in each tomographic bin), and for $\gamma_t$ there is also shape noise coming from the ellipticity dispersion $\sigma_{\epsilon}$  \cite{Crocce2011,Joachimi2008}. 

In general, we do not include the non-Gaussian covariance term in our analysis, as we are just interested in forecasting and comparing the cosmological constraints given by different sample definitions.
In addition, we exclude small scales (see Sec.~\ref{sec:scale-cuts}), where some of the non-Gaussian terms of the covariance become dominant (the  super-sample contribution also impacts large scales). We note that when comparing \redmagic with flux-limited samples, which have much higher number density, the latter will be more impacted by non-Gaussian terms due to the reduced shot noise in the Gaussian part of the covariance. Nonetheless, we have checked that including the non-Gaussian covariance term does not impact our final \maglim gains with respect to \redmagic after the optimization carried out in Sec.~\ref{subsec:maglim-optimization}.

We use two different codes to compute the Gaussian covariance: \cosmosis$ $ \cite{Zuntz2015} and  \cosmolike$ $ \cite{Cosmolike2017}, which was validated against simulations in \cite{Krause2017}. In Appendix~\ref{sec:cov-comp} we check that our results are the same independently of the code we use to compute the covariances.

\subsection{Parameter space and priors}
\label{sec:params-and-priors}

The cosmological model we consider in this work is spatially flat $w$CDM (i.e. having a free parameter for the dark energy equation of state) with fixed neutrino mass corresponding to the minimum allowed neutrino mass of 0.06 eV from oscillation experiments \cite{Patrignani2016}. We split the neutrino mass equally among the three eigenstates, to be consistent with \cite{DESY1_3x2}.
\begin{table}[H] 	
	\small
	\centering	
	\caption{\label{tab: priors}The fiducial parameter values and priors for cosmological and nuisance parameters used in this analysis. Square brackets denote a flat prior over the indicated range, while parentheses denote a Gaussian prior of the form $\mathcal{N}(\mu,\sigma)$.  \\} 
	\begin{tabular}{ccc}
		\hline
		\hline
		\textsc{Parameter} & \textsc{Fiducial} &\textsc{Prior}\\\hline
		\multicolumn{3}{c}{\textbf{\textsc{Cosmology}}}\\
		$\Omega_{\rm m}$ &  0.2837 &[0.1, 0.9] \\ 
		$A_\mathrm{s}/10^{-9}$ & 2.2606 & [$0.5$, $5.0$]  \\ 
		$n_{\rm s}$ & 0.9686 & [0.87, 1.07]  \\
		$w$ &  -1.0  &[-2, -0.33]   \\
		$\omb$ & 0.062 &[0.03, 0.07]  \\
		$h_0$  & 0.8433  &[0.55, 0.9]   \\
		$\Omega_\nu h^2$ & $6.155\times 10^{-4}$ &\textsc{Fixed} \\
		$\Omega_\mathrm{K}$ & 0 &\textsc{Fixed}\\ 
		$\tau$ & 0.08 &\textsc{Fixed}
		\\\hline
		
		\multicolumn{3}{c}{\textbf{\textsc{Galaxy bias} (\redmagic)} }	 \\
		$b^{i}$ & $1.4,1.6,1.6,1.93,1.99$ & [0.8,3.0]\\[0.15cm]\hline
		
		\multicolumn{3}{c}{\textbf{\textsc{Galaxy bias (MagLim)} } }	 \\
		$b^{i}$ & $1.49,1.86,1.81,1.90,2.26,2.33$ & [0.8,3.0]\\[0.15cm]\hline
		
		\multicolumn{3}{c}{\textbf{\textsc{Galaxy bias (Flux-limited)} } }	 \\
		$b^{i}$ & $1.07,1.24,1.34,1.56,1.96$ & [0.8,3.0]\\[0.15cm]\hline
		
		\multicolumn{3}{c}{\textbf{\textsc{Intrinsic alignment}} }	 \\
		$A_{\mathrm{IA}}$ & 0.0 & [-5.0,5.0]\\
		$\alpha_{\mathrm{IA}}$ & 0.0 & [-5.0,5.0]\\[0.15cm]\hline
		
		\multicolumn{3}{c}{\textbf{\textsc{ Lens photo-z shift (\redmagic)}}}	 \\
		$\Delta z_{{\rm l}}^{1}$ & 0.0 &(0.0,0.0035)\\
		$\Delta z_{{\rm l}}^{2}$ & 0.0  &
		(0.0,0.0035)\\
		$\Delta z_{{\rm l}}^{3}$ & 0.0 &
		(0.0,0.003) \\
		$\Delta z_{{\rm l}}^{4}$ & 0.0  &
		(0.0,0.005) \\
		$\Delta z_{{\rm l}}^{5}$ & 0.0 &
		(0.0,0.005) \\ [0.15cm]\hline
		
		\multicolumn{3}{c}{\textbf{\textsc{Lens photo-z shift (MagLim)}}}	 \\
		$\Delta z_{{\rm l}}^{1}$ & 0.0 &(0.0,0.007)\\
		$\Delta z_{{\rm l}}^{2}$ & 0.0  &
		(0.0,0.007)\\
		$\Delta z_{{\rm l}}^{3}$ & 0.0 &
		(0.0,0.006) \\
		$\Delta z_{{\rm l}}^{4}$ & 0.0  &
		(0.0,0.01) \\
		$\Delta z_{{\rm l}}^{5}$ & 0.0 &
		(0.0,0.01) \\
		$\Delta z_{{\rm l}}^{6}$ & 0.0 &
		(0.0,0.01)  \\[0.15cm]\hline
		
		\multicolumn{3}{c}{\textbf{\textsc{Lens photo-z shift (Flux-limited)}}}	 \\
		$\Delta z_{{\rm l}}^{1}$ & 0.0 &(0.0,0.014)\\
		$\Delta z_{{\rm l}}^{2}$ & 0.0  &
		(0.0,0.014)\\
		$\Delta z_{{\rm l}}^{3}$ & 0.0 &
		(0.0,0.012) \\
		$\Delta z_{{\rm l}}^{4}$ & 0.0  &
		(0.0,0.02) \\
		$\Delta z_{{\rm l}}^{5}$ & 0.0 &
		(0.0,0.02) \\[0.15cm]\hline
		
		\multicolumn{3}{c}{\textbf{\textsc{Source photo-z shift}} }	 \\
		$\Delta z_{{\rm s}}^{1}$ & 0.002 &(0.0,0.016)\\
		$\Delta z_{{\rm s}}^{2}$ & -0.015  &
		(0.0,0.013)\\
		$\Delta z_{{\rm s}}^{3}$ & 0.007 &
		(0.0,0.011) \\
		$\Delta z_{{\rm s}}^{4}$ & -0.018  &
		(0.0,0.022) \\[0.15cm]\hline
		
		\multicolumn{3}{c}{\textbf{\textsc{Shear calibration} }} 	 \\
		$m^{i}\,(i=1, 4)$ & 0.012 & (0.012, 0.023) \\[0.15cm]\hline
	\end{tabular}
\end{table}
The fiducial cosmological parameter values correspond to the best-fits of the posterior distributions from the DES Y1 \LCDM$\,$ analysis in \cite{DESY1_3x2} which obtained cosmological constraints from the combination of galaxy clustering, galaxy-galaxy lensing, and cosmic shear (a.k.a. 3$\times$2pt).

We bin the samples described in Sec.~\ref{sec:sample-selections} in several tomographic bins. For the $\maglim $ sample we split the selection in 6 redshift bins from $z=0.2$ to $z=1.05$, with a width of $\Delta z = 0.15$.  We consider the same $z$ range for the flux-limited sample, but in that case we split the selection in 5 $z$ bins with balanced number density across the bins. For $\redmagic $ we split the sample in 5 $z$ bins from $z=0.15$ to $z=0.95$, similarly to DES Y1 \cite{ElvinPoole2018}. See Table~\ref{tab:optimal-samples-stats} for the $z$ ranges in each tomographic bin of the samples. We keep fixed this fiducial redshift binning throughout this work, except for Sec.~\ref{sec:tomographic-binning} in which we consider alternative tomographic binnings.

\begin{figure}
	\begin{center}
		\includegraphics[width=\linewidth]{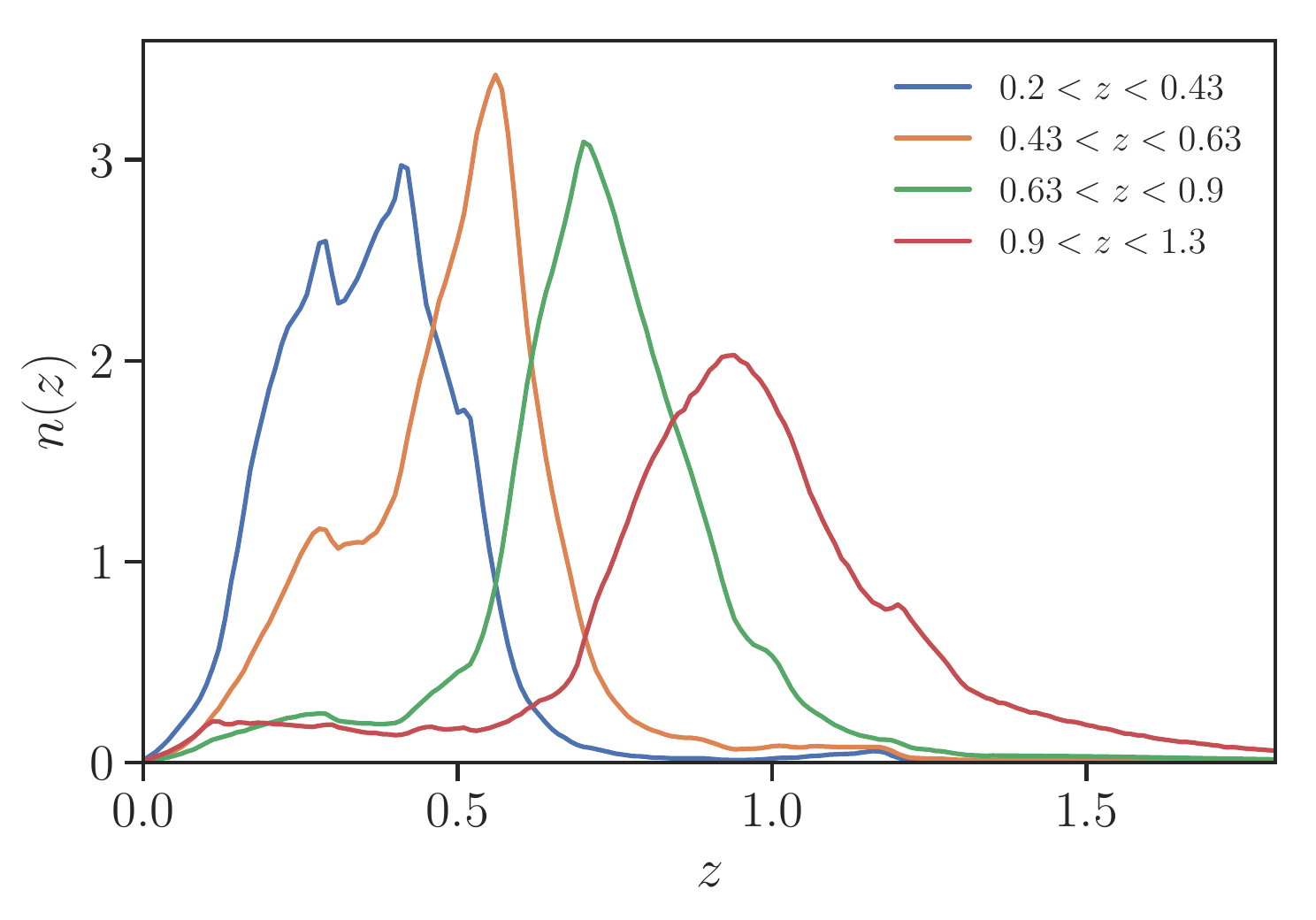}
		\caption{Normalized redshift distributions for the sources, corresponding to DES Y1 $\metacal$ galaxies.} 
		\label{fig:source}
	\end{center}
\end{figure}

For the sources we use the \textsc{Metacalibration} sample from the DES Y1 cosmic shear analysis \cite{Troxel2018}, which is divided in 4 tomographic bins: $0.2<z<0.43$, $0.43<z<0.63$, $0.63<z<0.9$, and $0.9<z<1.3$. See Figure~\ref{fig:source} for the normalized redshift distributions.

In addition to the 6 cosmological parameters, our model contains about 20 nuisance parameters (22 for $\maglim $ due to the extra redshift bin). These are the galaxy bias parameters  for the lens samples (one $b^i$ per redshift bin), the multiplicative shear biases (one $m^i$ parameter for each source redshift bin), two parameters related to the intrinsic alignment model, $A_{\mathrm{IA}}$ and $\eta_{\mathrm{IA}}$, and the photo-z shift parameters for each redshift bin of the lenses and the sources, $\Delta z^i$. 

These shift parameters are used in our analysis to quantify uncertainties in the redshift distribution. We assume that the true redshift distribution $n^i(z)$ in bin $i$ is a shifted version of the photometrically derived distribution:
\begin{equation}
n^i(z)=n^i_{\mathrm{PZ}}(z-\Delta z^i).
\end{equation}
The fiducial values and priors assumed for these parameters, shown in Table \ref{tab: priors}, are consistent with the DES Y1 3$\times$2 analysis \cite{DESY1_3x2}, except that the lens photo-z shifts are treated as described below. For the $\maglim $ sample, we assume fiducial values for the galaxy bias based on galaxy clustering measurements on a 10\% subsample of the data, in consistency with the Y3 blinding scheme \cite{y3-blinding}. 

For the flux-limited sample we assume fiducial galaxy bias values based on the galaxy clustering measurements from DES Science Verification (SV) data \cite{Crocce2016}, where a similar flux-limited sample was defined. In Sec.~\ref{sec:galaxy-bias} we check that our conclusions in this work are basically insensitive to changes in the fiducial galaxy bias values.

For the \photoz shift parameters we assume the same priors as in DES Y1 for the sources, since we are using the same redshift distributions. 
For the lenses, in the DES Y1 data analysis the shift values and their associated errors were obtained by re-calibrating the mean of the baseline redshift distributions to match those from a clustering-redshift method, given a reference spectroscopic sample. In DES Y1 this sample was made of $\sim 20,000$ CMASS and LOWZ galaxies in $\sim$ 124 $\sqdeg$ area overlap with SDSS DR12 \cite{2018MNRAS.481.2427C}. For the Y3 analysis the DES footprint overlaps over a much larger area with SDSS DR12 in addition to eBOSS, which increases the reference sample by about a factor of ten in number of galaxies \citep[][in prep.]{Cawthon2020}. Hence the associated errors $\sigma$ are found a factor of $\sim$ 2 smaller for \redmagic than in Y1 \cite{Cawthon2020}.  In turn, \maglim has broader redshift distributions than \redmagic and the errors on the shift parameters from the clustering-redshift method in Y3 are roughly twice as big than for \redmagic. Similarly, since the flux-limited sample has even broader redshift distributions (see Figures~\ref{fig:sigz} and \ref{fig:nz-optimal-samples}), we conservatively assume priors twice as wide compared to $\maglim$, which is a reasonable assumption according to Y3 clustering-redshift estimates \cite{Cawthon2020}. In Sec.~\ref{sec:photoz-uncertainties} we test the sensitivity of our results to the assumed priors for the $\maglim$ and flux-limited lens \photoz shift parameters.

\subsection{Scale cuts}
\label{sec:scale-cuts}

 At sufficiently large scales, perturbation theory can be used to calculate the matter power spectra. On smaller scales, N-body simulations are needed in order to capture the non-linear evolution of structure growth. For example, the \textsc{Halofit} method \cite{Smith2003,Takahashi2012}, which we use in this work, employs a functional form of the matter power spectrum derived from halo models that are, in turn, calibrated from N-body simulations. However, only gravitational physics is included in these dark matter only simulations, which neglects any modification of the matter distribution due to baryonic physics processes such as star formation, radiative cooling, and feedback \cite{Cui2014,Velliscig2014,Mummery2017}. At small scales, these processes can modify the matter power spectrum significantly \cite{vanDaalen2011}.
 
In order to mitigate the impact of the uncertainty in how the baryonic physics and other non-linear effects modify the matter power spectrum, we apply a set of scale cuts, which were tested in \cite{Krause2017} for the DES Y1 analysis, such that non-linear modeling limitations (especially in the galaxy bias modeling) do not bias the cosmology results. In this work we use the same scale cuts considered for the DES Y1 baseline analysis \cite{DESY1_3x2}, which are defined in terms of a specific comoving scale $R$,
 \begin{equation}
 \begin{gathered}
 R_{\delta_g \delta_g} =8 \, \Mpc \, h^{-1},\\
 R_{\delta_g \kappa} =12 \, \Mpc\, h^{-1},
 \end{gathered}
 \end{equation}
 where $R_{\delta_g \delta_g}$ denotes the scale cuts for the galaxy clustering data vector, and $R_{\delta_g \kappa}$ for galaxy-galaxy lensing. 
 See \cite{Krause2017} for a detailed description of how these scale cuts were determined. We then convert the comoving scale cuts into angular ones using the radial comoving distance $\chi$ to the mean of the redshift distribution in each corresponding tomographic bin $\left\langle z^i \right\rangle$. Thus, for redshift bin $i$ the minimum angular scale $\theta^i_{\min}$ included is, 
 \begin{equation}
 \theta^i_{\min} = \dfrac{R}{\chi\,(\left\langle z^i \right\rangle)}.
 \end{equation}

\section{Sample Optimization}
\label{sec:sampleoptimization}

In this section, we explore the trade-off between number density and \photoz scatter by considering different flux-limited sample definitions. In particular, we define different selections for the samples described in  Sec.~\ref{sec:flux-limited-sample} and Sec.~\ref{sec:maglim-sample} and see how that impacts the constraints on $w$, $\sigma_8$ and $\Omega_m$.  We fix the fiducial galaxy bias, tomographic binning, and nuisance parameters as specified in Sec.~\ref{sec:params-and-priors}. The impact of fixing these is discussed in Sec.~\ref{sec:analysis-choices}, in which we show that our conclusions are robust to the galaxy bias and tomographic binning assumed. We consider an area of 4580 $\sqdeg$ for all the forecasts in this paper, even though this value is different to the final area of the data catalog, which was reduced after masking (see Sec.~\ref{sec:survey-area}). For each one of the galaxy selections, we only vary the photometric redshift distribution of the lens sample and its tomographic number densities. In all cases, we use the DES Y1 \metacal sample for the sources.



\subsection{$\maglim $ sample}
\label{subsec:maglim-optimization}

As presented in Sec.~\ref{sec:maglim-sample}, we consider samples in which all galaxies have a magnitude cut applied that evolves linearly with the photometric redshift estimate: $i < a z_{\rm phot} + b$. In this section we consider different values of $a$ and $b$, in a range wide enough to cover a variety of number densities and $\sigma_z$ values.

\begin{figure}
	\begin{center}
		\includegraphics[width=\linewidth]{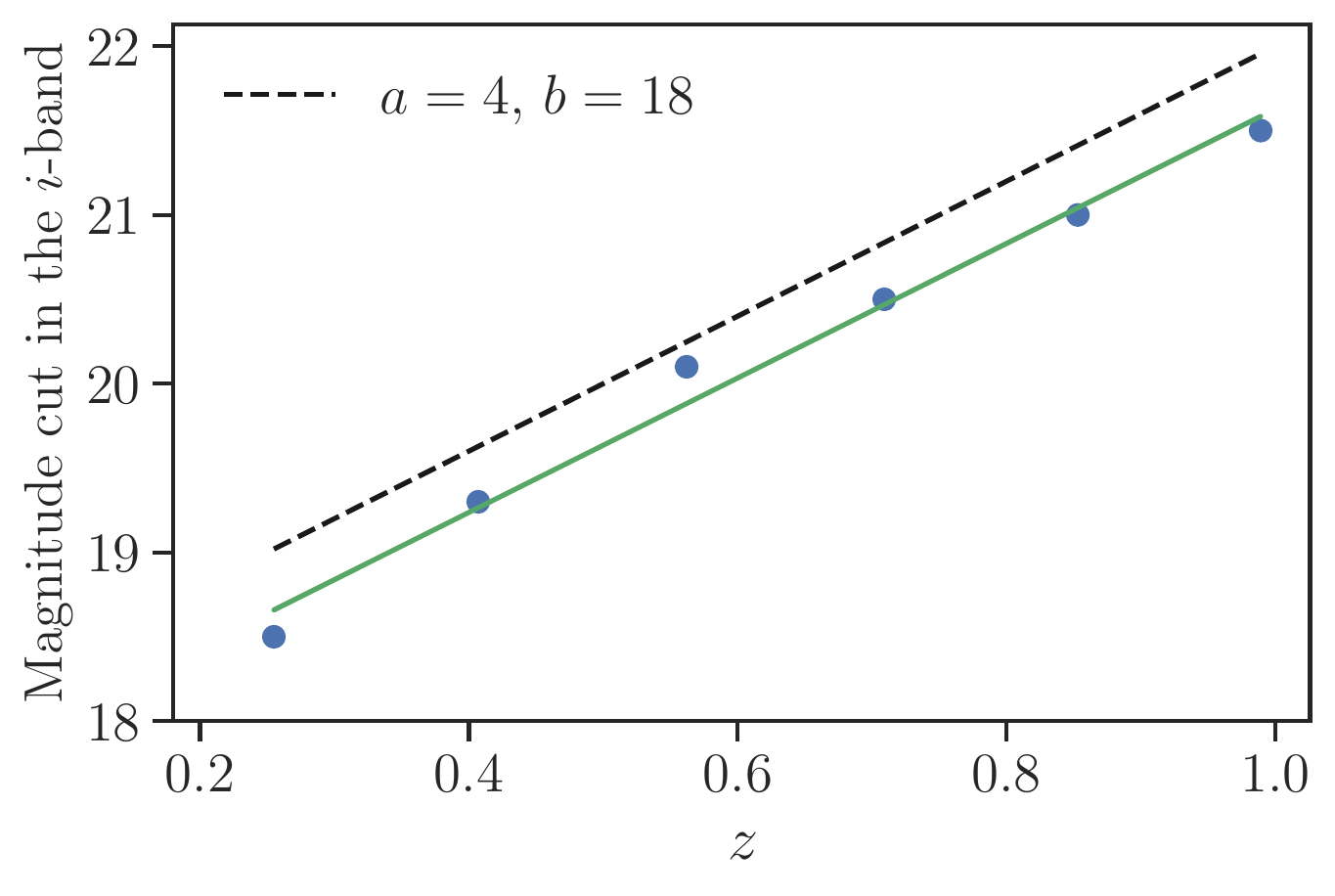}
		\caption{Different $\maglim$ sample definitions considered. The first version (blue dots) applied a constant magnitude cut for each redshift bin, the second version (a.k.a. v0.0), in solid green, used a continuous magnitude cut evolving linearly with $z$, with slope and interception given by a fit to the blue points. In dashed black we show the final definition of the sample.}
		\label{fig: mag_cut}
	\end{center}
\end{figure}

In order to get a first estimate for these values, we start by applying a different limiting magnitude in the $i$ band to each redshift bin, aiming for a number density two to three times larger than $\redmagic $ while keeping the photo-z scatter as low as possible. The resulting limiting magnitudes are shown in Figure~\ref{fig: mag_cut} (blue points). We then fit the linear function to these $i$ and $z$ values obtaining $a=4.0$ and $b=17.64$. In Figure~\ref{fig: mag_cut} we show the $i$ values used for the preliminary version of the sample, the linear fit to these values (green), hereafter v0.0, and the cut corresponding to the optimal definition of the sample (see Sec.~\ref{sec:sample-comparison}). In order to find the optimal sample we follow these steps:
\begin{enumerate}
	\item Take one of the possible combinations of $(a,\ b)$ within the ranges $a=[3.5, 4, 4.5, 5]$, $b=[17, 17.5, 18, 18.5]$.
	\item Apply the cut $i < a z_{\rm phot} + b$ with the selected $a$ and $b$ values.
	\item From this selection we extract the redshift distributions $n(z)$ and number densities, which will be used as input for the forecasts.
	\item Generate a covariance and a theory data vector using as input for the lenses the $n(z)$ for this sample selection (and the number densities, in the case of the covariance).
	\item Using this theory data vector and covariance, we run a 2$\times$2pt Fisher forecast to obtain estimated constraints and FoM on the parameters of interest (see Table~\ref{tab: priors}). 
\end{enumerate}
 As mentioned before, these ranges of $(a,\ b)$ values cover a broad variety of possible sample definitions, as the minimum values (i.e. $i < 3.5z_{\rm phot} + 17$) result in a sample with very few galaxies (about 75 galaxies per $\sqdeg$), and the maximum ones (i.e. $i < 5z_{\rm phot} + 18.5$) result in a sample with a very large limiting magnitude ($i < 23.75$), in such a way that we are practically selecting almost all the galaxies from the catalog (roughly 15300 galaxies per $\sqdeg$). As discussed in Sec.~\ref{sec:survey-area}, we decide not to consider those selections that reach a limiting magnitude larger than 22.75, at which we already lose $\sim10\%$ of the area (see Figure~\ref{fig:area}). 

\begin{figure}
	\begin{center}
		\includegraphics[width=\linewidth]{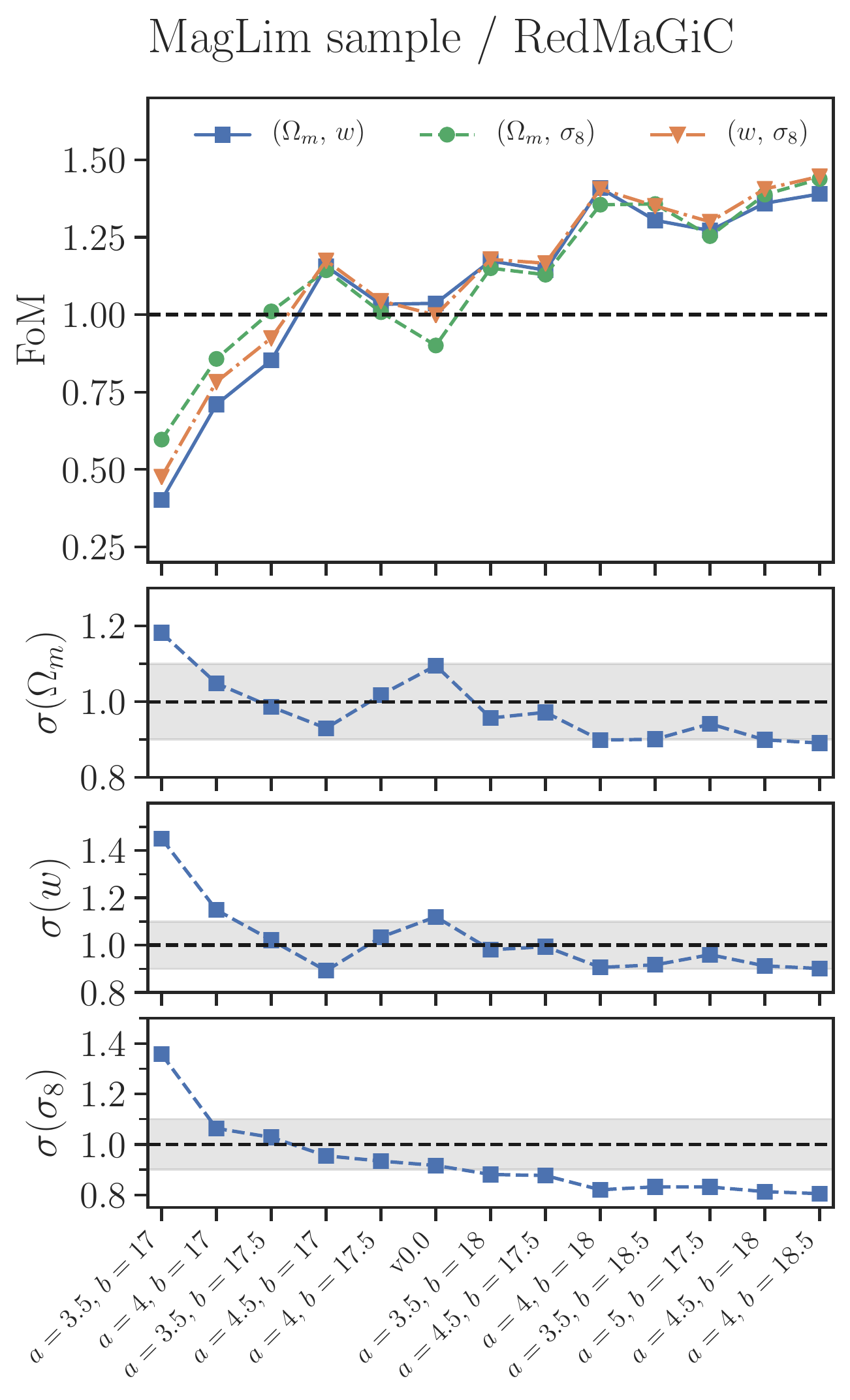}
		\caption{Standard deviations on $\Omega_m$, $w$ and $\sigma_8$ (bottom panel) and the figure of merit of their combinations in pairs (top panel)  considering different magnitude limited samples (of the form $i < a z_{\rm phot} + b$) normalized by estimates from the \redmagic$ $ sample. The gray band delimits the region with $10\% $ better (lower edge) or worse (upper edge) constraints compared to \redmagic. The samples are ordered by ascending number density (from left to right), with values ranging from $\sim75$ to $\sim5775$ galaxies per $\sqdeg$.}
		\label{fig: fisher_comp}
	\end{center}
\end{figure}

\begin{figure}
	\begin{center}
		\includegraphics[width=\linewidth]{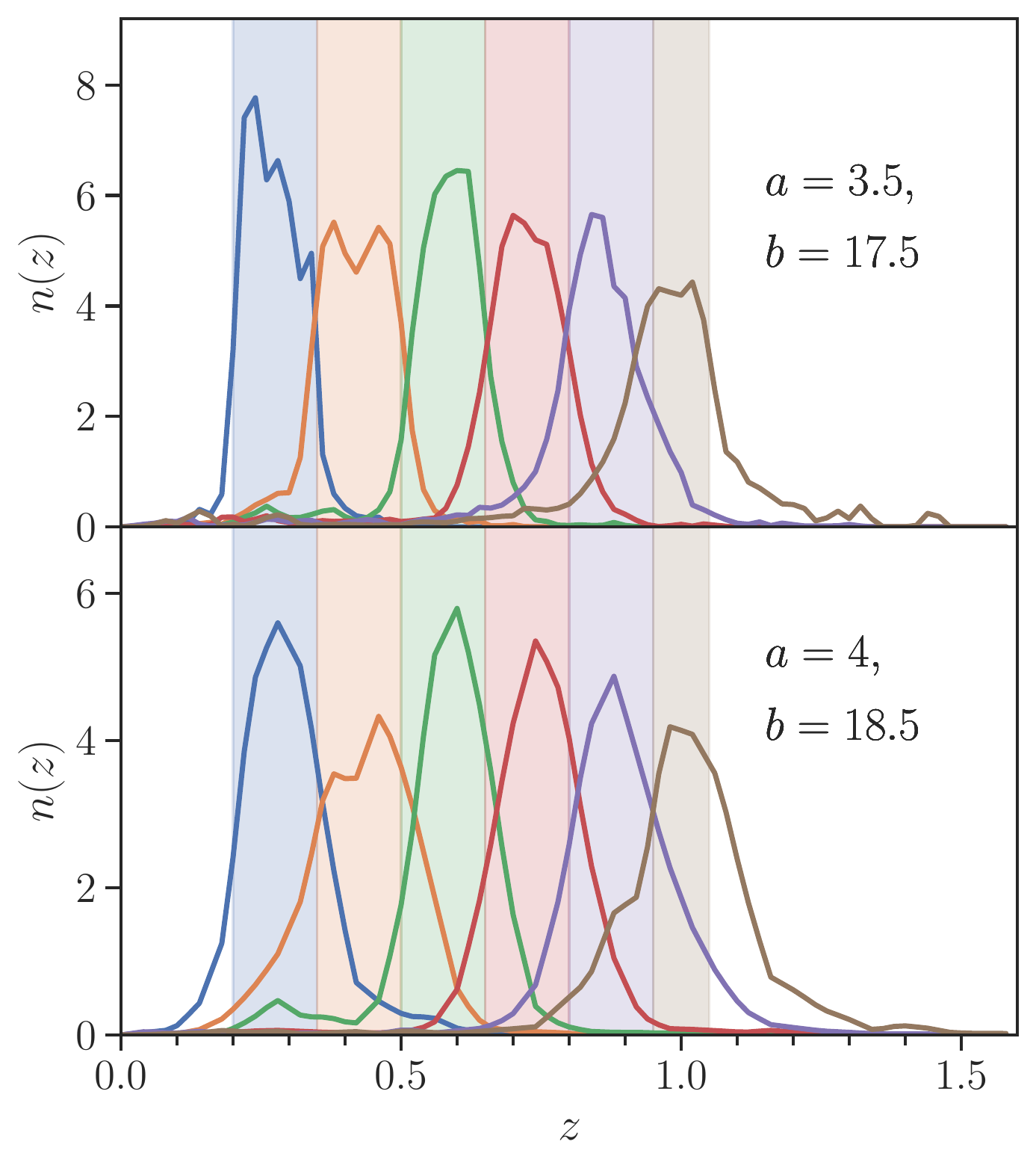}
		\caption{Normalized redshift distributions for two magnitude-limited sample selections with significantly small (top panel) and large (bottom panel) number densities (see Figure~\ref{fig: fisher_comp}). The mean photo-z scatter ranges from $\sigma_z/(1+z) \approx 0.028$ in the top panel to  $\sigma_z/(1+z) \approx 0.050$ in the bottom panel. The shaded bands indicate the tomographic binning assumed.  }
		\label{fig: maglim optim nz comp}
	\end{center}
\end{figure}


 In the bottom three panels of Figure~\ref{fig: fisher_comp}, we show the standard deviations resulting from the forecasts, which are normalized by the constraints obtained from the \redmagic sample. Thus, the black dashed line represents constraints equal to those obtained from \redmagic, while points above or below that line correspond to samples giving worse or better constraints than \redmagic, respectively. The grey band delimits the region with 10\% better or worse constraints.
In the top panel we show the respective figure of merits for each pair combination of these cosmological parameters, also normalized by the FoM obtained with $\redmagic $. Note that tighter constraints imply larger FoM values. 

Here we see that most of the samples considered yield constraints similar or slightly better than \redmagic. This is due to the fact that, even though the photo-z are less accurate, these samples have more galaxies and reach higher $z$ than  \redmagic (recall we consider $z_{\max} =  1.05$, while for \redmagic $z_{\max}=0.95$). One of the samples provides significantly worse constraints ($i < 3.5z_{\rm phot} + 17$), but this is  understandable, as it corresponds to the extreme case in which very few galaxies are selected from the data catalog.

It is interesting to note that the constraints on $\sigma_8$ improve as the number density increases. For $\Omega_m$ and $w$ this trend is not so clear, in part due to the trade-off with photometric redshift accuracy which widens the redshift distributions as the number density increases. This trade-off can be seen more clearly in Figure~\ref{fig: maglim optim nz comp}, in which we compare the normalized redshift distributions of two magnitude-limited sample selections ordered by ascending number density (and, consequently, mean \photoz scatter) from top to bottom. These correspond to sample selections from Figure~\ref{fig: fisher_comp} with significantly small and large number densities. 

 Another factor to take into account is that different combinations of $a$ and $b$ in the selection $i<az_{\rm phot}+b$ result in uneven distributions of number densities across the tomographic bins. Since we are comparing the constraints from the joint combination of galaxy clustering and galaxy-galaxy lensing, we expect to have an increased constraining power from those samples that have more galaxies at the redshifts in which the lensing efficiency kernels of the source sample peak. Thus, a sample that has more galaxies at high redshift and fewer galaxies at low redshift can provide tighter constraints than a sample with the same total number density but with the opposite distribution of galaxies.

From Figure~\ref{fig: fisher_comp} we see that the optimal sample, i.e. the one that produces the tightest constraints (higher FoM) while keeping the \photoz uncertainties as low as possible, corresponds to $i<4 z_{\rm phot} + 18$. 
With a number density $2-3$ times larger than $\redmagic$ (see Sec.~\ref{sec:sample-comparison}), this sample has an increase in the FoM values of 40$\%$ (36$\%$ for the $\Omega_m - \sigma_8$ pair), providing $\sim10-18\%$ smaller errors on the cosmological parameters.

We note that the sample with the largest number density from Figure~\ref{fig: fisher_comp}, $i<4 z_{\rm phot} + 18.5$, provides very similar constraints to $i<4 z_{\rm phot} + 18$. However, this sample has larger mean \photoz scatter $\sigma_z$ in all tomographic bins. Aside from increasing the width of the redshift distributions (see Figure~\ref{fig: maglim optim nz comp}), this could present more obstacles in the validation of the redshift distributions in a real data analysis. For this reason, the selection $i<4 z_{\rm phot} + 18$ is preferable.



\subsection{Flux-limited sample}

\begin{figure}
	\begin{center}
		\includegraphics[width=\linewidth]{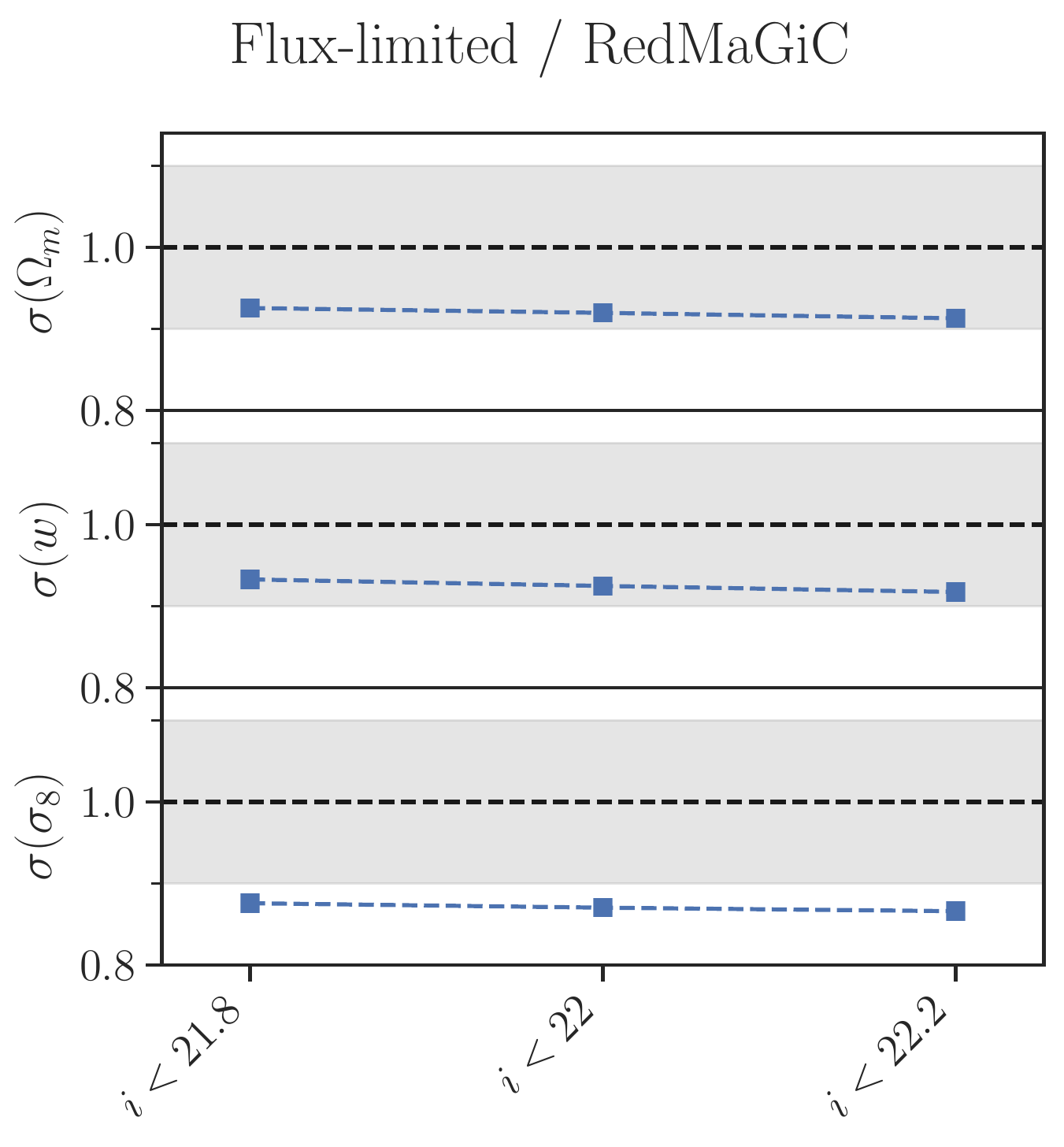}
		\caption{Standard deviations on $\Omega_m$, $w$ and $\sigma_8$  considering different definitions of the flux-limited sample, normalized by estimates from the \redmagic$ $ sample. The gray band delimits the region with $10\% $ better (lower edge) or worse (upper edge) constraints compared to \redmagic. }
		\label{fig:fluxlim-optimization}
	\end{center}
\end{figure}


In this section we explore flux-limited samples with different limiting magnitudes in the $i$ band, as described in Sec.~\ref{sec:flux-limited-sample}. We restrict ourselves to a maximum limiting magnitude of 22.2 to avoid having to mask out a larger fraction from our total area (see Sec.~\ref{sec:survey-area}), this also corresponds to the limiting magnitude of the optimal $\maglim $ sample ($i < 4 z_{\rm phot} + 18$).

Similarly to the optimization process described for \maglim in the previous section, we run $2\times$2pt Fisher forecasts for three limiting magnitudes: [21.8, 22, 22.2]. In Figure~\ref{fig:fluxlim-optimization} we compare the constraints obtained on $\Omega_m$, $w$, and $\sigma_8$ for each one of these flux-limited samples normalized by the $\redmagic $ ones. The shaded grey band delimits the region with 10\% worse or better constraints compared to \redmagic. Even though there is a significant variation in number densities in the samples considered (e.g. $i<22.2$ has twice the number density of $i<21.8$ at $0.8 < z < 1.05$), there is not much difference in the resulting constraints. However, we note that the scale cuts considered are conservative, and the difference in constraining power would be larger if we included smaller scales in our analysis.  We find a slight improvement when increasing the number densities (going to higher limiting magnitudes). Thus, the optimal flux-limited sample is the one with limiting magnitude $i<22.2$, with which we reach ~8\% tighter constraints on $\Omega_m$ and $w$, and $13\%$ tighter on $\sigma_8$ with respect to $\redmagic$.  These improvements would be likely smaller when including shear, i.e. in the usual $3\times2$pt analysis.

Comparing these results to those obtained for $i<4 z_{\rm phot} + 18$ (\maglim) we find that these constraints are somewhat worse, especially for $\sigma_8$.
The reason for this  is the trade-off between number density and photo-z scatter. The flux-limited samples have much higher number density than  $\maglim $ (see Figure~\ref{fig:sigz}), which in general improves the constraints because it reduces the shot noise contribution in the covariance. But at the same time, the larger $\sigma_z$ increases the errors on the cosmological parameters, partly due to the wider priors in the lens photo-z shift parameters $\Delta z^i$, and partly because the redshift distributions have larger tails and we are not including galaxy clustering cross-correlations between redshift bins. 
Moreover, due to the larger uncertainties in the shapes of the redshift distributions, it is not clear that a \photoz shift parameter is enough to account for these uncertainties. The addition of extra nuisance parameters (e.g. a \photoz width parameter for each bin) may be needed in a real data analysis, and this could degrade the constraining power of this sample.
Nevertheless, we note that exploring smaller scales will be more beneficial for \maglim and, specially, for the  flux-limited sample, as they are on the sample variance regime (while \redmagic is shot noise limited).

\section{Optimal Samples}
\label{sec:sample-comparison}

\setlength{\tabcolsep}{10pt} 
\begin{table}
	\centering
	\caption{\label{tab:optimal-samples-stats} Number of galaxies, mean photo-$z$ scatter, and  68\% confidence width of the redshift distributions ($W_{68}$)  for the optimal $\maglim $ and flux-limited samples compared to \redmagic, considering an effective area of 4182 $\sqdeg$. }
	\begin{tabular}{cccc}
		\hline
		\hline
		\textsc{$z$ range} & $n_{\delta_g}$  & $\sigma_z/(1+z)$ & $W_{68}$\Tstrut \\\hline
		\multicolumn{4}{c}{\textbf{\redmagic}}\Tstrut\\
		
		0.15 -- 0.35 & 341,602 & 0.011 & 0.059\Tstrut\\
		0.35 -- 0.50 & 589,562 & 0.015 & 0.052\\
		0.50 -- 0.65 & 877,267 & 0.016 & 0.052\\
		0.65 -- 0.85 & 679,291 & 0.020 & 0.073\\
		0.85 -- 0.95 & 418,986 & 0.022 & 0.050\Bstrut   \\  \hline
		
		\multicolumn{4}{c}{\textbf{$\maglim$}}\Tstrut \\
		
		0.20 -- 0.35 & 1,680,160 & 0.034 & 0.064\Tstrut \\
		0.35 -- 0.50 & 1,678,655 & 0.043 & 0.082\\
		0.50 -- 0.65 & 1,460,354 & 0.022 & 0.061\\
		0.65 -- 0.80 & 1,975,242 & 0.027 & 0.069\\
		0.80 -- 0.95 & 2,374,205 & 0.034 & 0.077\\
		0.95 -- 1.05 & 1,470,893 &  0.044 & 0.097\Bstrut\\\hline
		
		\multicolumn{4}{c}{\textbf{Flux-limited}}\Tstrut\\
		
		0.20 -- 0.40 &  12,623,785 & 0.061 & 0.113\Tstrut\\
		0.40 -- 0.50 & 16,291,232 	& 0.066 & 0.101\\
		0.50 -- 0.65 & 16,795,581 & 0.050 & 0.098\\
		0.65 -- 0.80 & 12,994,143 & 0.036 & 0.077\\
		0.80 -- 1.05 & 11,244,729 & 0.040 & 0.110\Bstrut\\\hline
		
	\end{tabular}
\end{table}

\begin{figure}
\begin{center}
\includegraphics[width=\linewidth]{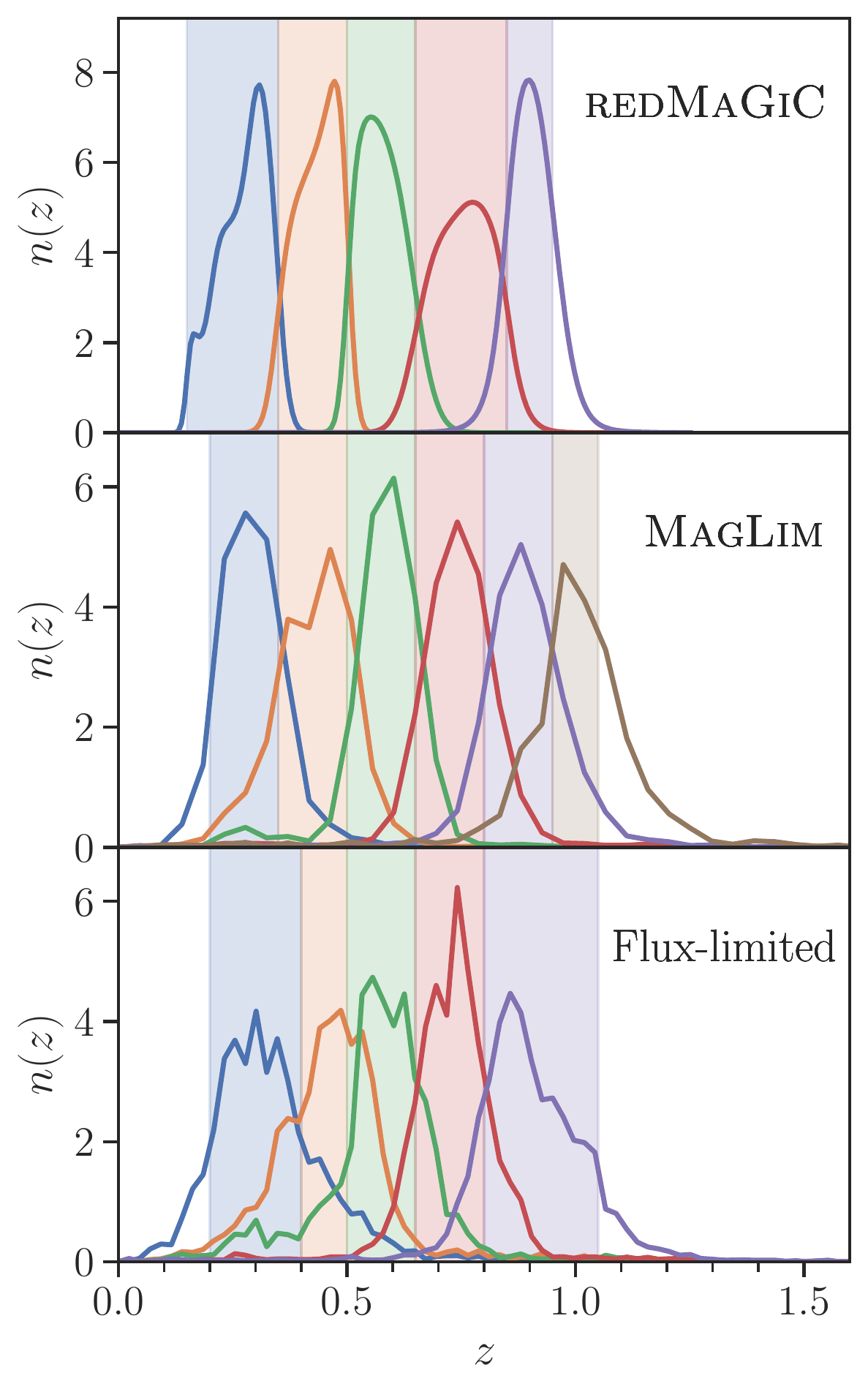}
\caption{Normalized redshift distributions for  the optimal $\maglim$ and flux-limited samples compared to \redmagic. The shaded bands indicate the tomographic binning of each sample. }
\label{fig:nz-optimal-samples}
\end{center}
\end{figure} 

In Sec.~\ref{sec:sampleoptimization} we find that the optimal sample is the $\maglim $ sample, defined with a magnitude cut $i< 4 z_{\rm phot}  +18$. In Table~\ref{tab:optimal-samples-stats} we describe the fiducial tomographic binnings of the three samples, along with the number of galaxies, $n_{\delta_g}$, the mean photo-z scatter, $\sigma_z/(1+z)$, and the 68\% confidence width of the redshift distributions $W_{68}$ in each redshift bin. The $W_{68}$ value is the equivalent of the standard deviation of a Gaussian distribution, and in practice is much more relevant to consider than $\sigma_z$ itself because it is a measure of the width of the redshift distribution, which is what enters the density kernel in the two-point functions computation (see Sec.~\ref{sec:theory}). 
In Figure~\ref{fig:nz-optimal-samples} we show the redshift distributions for the three samples. The flux-limited sample is the one with larger photo-z scatter, and as a consequence, the redshift distributions are broader than with the other two samples. 

In order to compare the properties of the samples under the same terms, in the following Sec.~\ref{sec:comp-same-zbinning} we compare the number of galaxies and $W_{68}$ values assuming the same tomographic binning. In Sec.~\ref{subsec:cosmological-constraints} we compare their cosmological constraints obtained from full MCMC simulated likelihood analyses.

\subsection{Comparison on same tomographic binning}
\label{sec:comp-same-zbinning}

In this section we compare the characteristics (number density and photometric accuracy) of $\maglim $ with the flux-limited and $\redmagic $ samples under the same tomographic binning. In particular, we assume the $\redmagic $ tomographic binning for the three samples. Since the $\maglim $ and flux-limited samples reach a higher maximum redshift than $\redmagic $, for those two samples we consider an additional redshift bin in the range $0.95<z<1.05$.

In Table~\ref{tab:comp-same-zbinning} we show the  number of galaxies and $W_{68}$ values for each redshift bin and each one of the lens samples.  The $\maglim $ sample has on average between 2 and 3 times more galaxies than \redmagic. The difference in number density ranges from 60\%  more galaxies in the third bin to more than 5 times further galaxies at higher redshift ($0.85<z<0.95$), while the redshift distributions are $\sim30\%$ wider on average for the $\maglim $ sample.

On the other hand, the number of galaxies in the flux-limited sample is one order of magnitude larger compared to $\maglim $, except at high redshift ($0.85<z<1.05$) where the $\maglim $ selection gets closer to the flux-limited selection of $i<22.2$, as they both have the same limiting magnitude at $z_{\max} = 1.05$. The flux-limited sample has a high number density at the expense of larger \photoz errors (see Figure~\ref{fig:sigz}). As a consequence, its redshift distributions are on average $~20\%$ wider compared to $\maglim$, with the difference being larger in the range $0.50 < z < 0.65$. 

The greater number density of $\maglim$ and flux-limited samples compared to $\redmagic$ is the dominant factor driving the gain of constraining power from a 2x2pt analysis. The extension to higher redshift ($z_{\max}=1.05$) is a sub-dominant effect in this case due to the weak lensing kernels peaking at $z\sim0.6$ (see Figure~\ref{fig:source}). Therefore, the increase in number density in the other tomographic bins ($z<0.95$) dominates the overall gain of these samples compared to $\redmagic$.

\begin{table}
	\centering
	\caption{\label{tab:comp-same-zbinning} Comparison of number of galaxies and  68\% confidence width of the redshift distribution, $W_{68}$, for the optimal $\maglim $ and flux-limited samples compared to \redmagic, considering an effective area of 4182 $\sqdeg$ and the same tomographic binning.}
	\begin{tabular}{lccc}
		\hline
		\hline
		$\,$ \textsc{$z$ range}  & \redmagic & $\maglim$ &  \textsc{Flux-lim}\Tstrut \\\hline
			 &\multicolumn{3}{c}{\textbf{Number of galaxies}}\Tstrut \\ 
		0.15 -- 0.35 & 341,602 & 1,599,462 & 9,129,473\Tstrut \\
		0.35 -- 0.50 & 589,562 & 1,593,745 &21,473,232\\
		0.50 -- 0.65 & 877,267 & 1,379,717 &16,795,581\\
		0.65 -- 0.85 & 679,291 & 1,862,978 & 16,640,513\\
		0.85 -- 0.95 & 418,986 & 2,257,704 & 5,093,174 \\
		0.95 -- 1.05 & 				& 1,470,893 &  2,503,679\Bstrut \\\hline
		$\, \, \, \,$ \textsc{Total} & 2,906,708  & 10,164,499 & 71,635,652\Tstrut\\\hline
		
		& \multicolumn{3}{c}{\textbf{Width of the redshift distribution ($W_{68}$)}}\Tstrut\\ 
		0.15 -- 0.35 & 0.059 &  0.073  & 0.088\Tstrut\\
		0.35 -- 0.50 & 0.052  & 0.082	& 0.105 \\
		0.50 -- 0.65 &0.052 & 0.061 & 0.098 \\
		0.65 -- 0.85 &0.073 & 0.085 & 0.091\\
		0.85 -- 0.95 & 0.050 & 0.076 & 0.086\\
		0.95 -- 1.05 &  & 0.097 &  0.096\Bstrut\\\hline
		
	\end{tabular}
\end{table}

\subsection{Cosmological constraints from MCMC likelihood analysis}
\label{subsec:cosmological-constraints}

\begin{figure}
	\begin{center}
		\includegraphics[width=\linewidth]{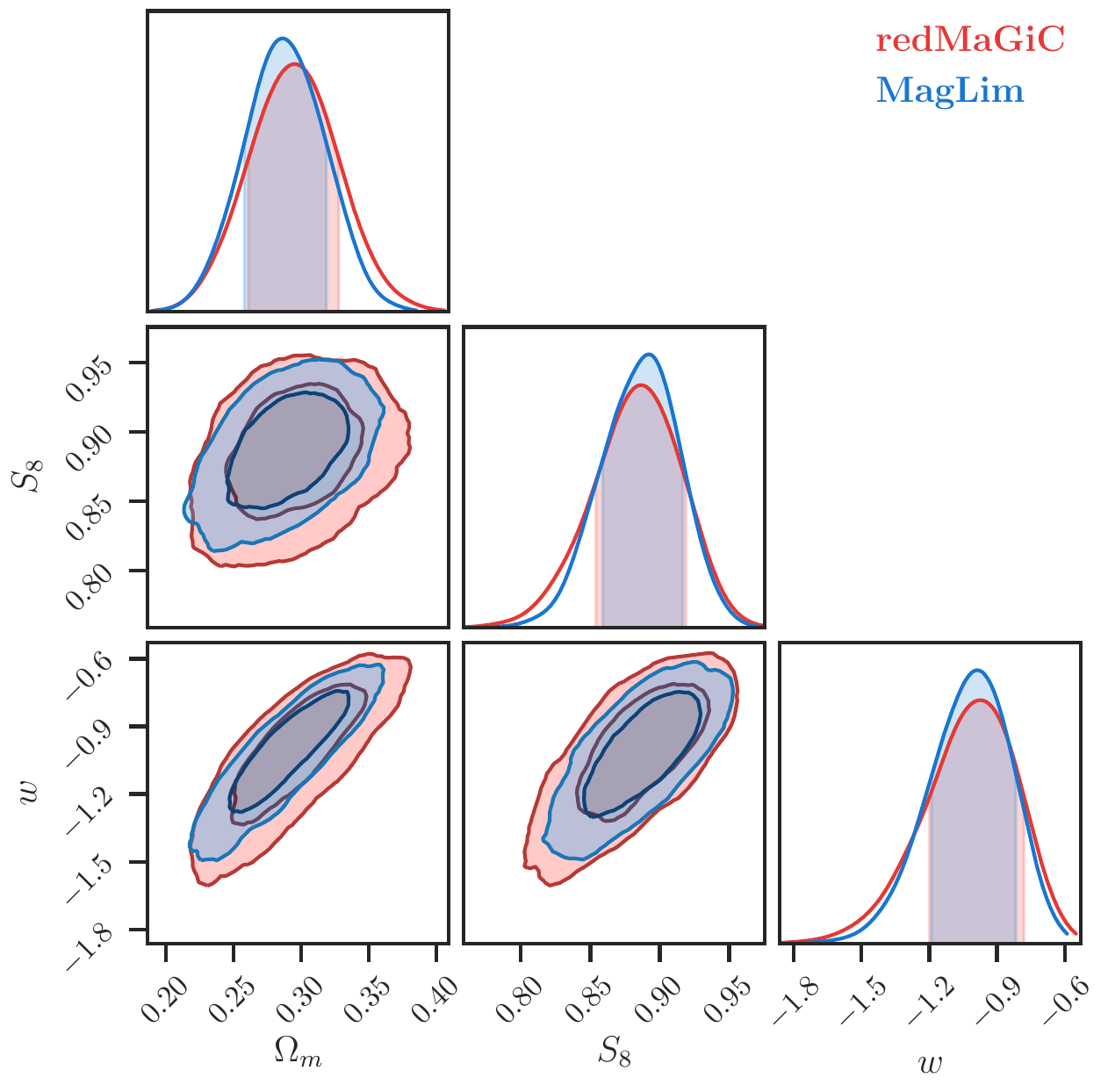}
		\caption{Comparison of 2$\times$2pt parameter constraints obtained using the DES Y3 \redmagic (red) and $\maglim $ (blue) samples and the DES Y1 \metacal source sample. Here, and in all the 2D plots below, the two sets of contours depict the 68\% and 95\% confidence levels (CL). The $\maglim$ constraints  are tighter by 10\% on $\Omega_m$, 13\% on $S_8$, and 12\% on $w$ compared to $\redmagic$.}
		\label{fig:wcdm-mcmc}
	\end{center}
\end{figure}

In this section, we compare the cosmological constraints obtained from the optimal $\maglim$ and flux-limited samples with respect to the \redmagic sample after performing a full MCMC analysis of the combination of galaxy clustering and galaxy-galaxy lensing, as opposed to the Fisher matrix approach taken in the other sections. We assume the fiducial values and priors listed in Table~\ref{tab: priors} and the tomographic binnings from Table~\ref{tab:optimal-samples-stats}. However, in addition to exploring the constraints on $\sigma_8$, in this section we also consider   the related parameter
 \begin{equation}
 S_8 \equiv \sigma_8 \left( \dfrac{\Omega_m}{0.3} \right)^{0.5},
 \label{eq:S8}
 \end{equation}
since $S_8$ is better constrained than $\sigma_8$ in weak lensing surveys such as DES, and it is largely uncorrelated with $\Omega_m$ in the DES parameter posterior.

In Figure~\ref{fig:wcdm-mcmc} and Table~\ref{tab:mcmc-constraints} we show the constraints on $\Omega_m$, $S_8$, and $w$ ($\sigma_8$ constraints also included in Table~\ref{tab:mcmc-constraints}). We find that, using the $\maglim $ sample instead of \redmagic, we obtain  10\% tighter constraints on $\Omega_m$,  about 12-13\% for $S_8$ and $w$, and  $16\%$ on $\sigma_8$ .  Regarding the flux-limited sample, we generally find worse constraints compared to $\maglim$, with the difference being  2\% on $\Omega_m$, and  6-7\% on $w$ and $\sigma_8$. However, when sampling the $S_8$ parameter, the flux-limited sample provides an 11\% improvement with respect to $\maglim$.  This is due to the flux-limited sample having a projected 2D posterior in the $S_8 - \Omega_m$ plane with a slightly different inclination compared to \maglim, favoring tighter $S_8$ constraints.

\begin{figure}
	\begin{center}
		\includegraphics[width=\linewidth]{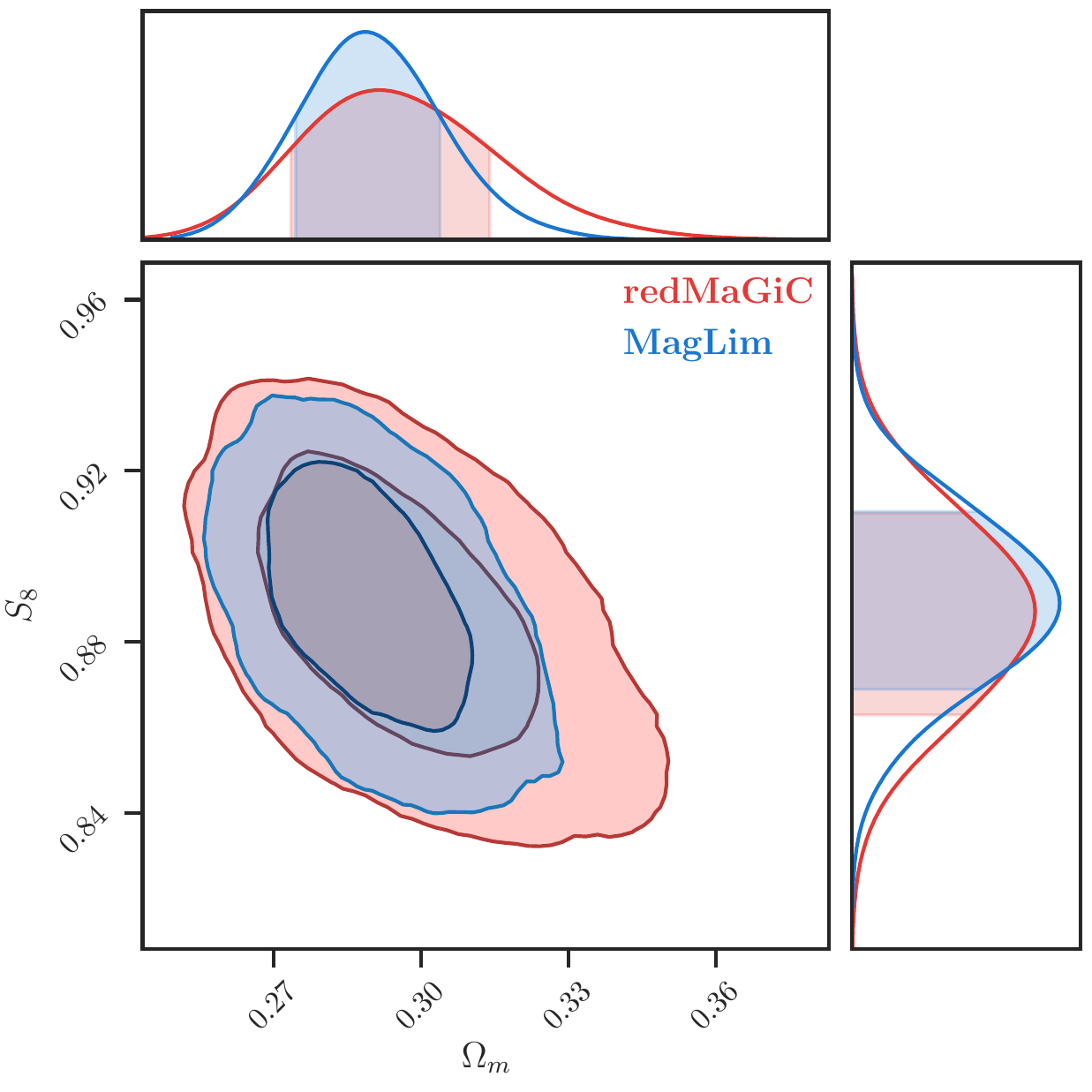}
		\caption{\LCDM$ $ 2$\times$2pt constraints (fixing $w$) using the DES Y3 \redmagic (red) and $\maglim $ (blue) samples as lenses and the DES Y1 \metacal sample  as sources. The $\maglim $ constraints are tighter by 27\% on $\Omega_m$, and 11\% on $S_8$ compared to $\redmagic$.}
		\label{fig:lcdm-mcmc}
	\end{center}
\end{figure}

\begin{table}
	\centering
	\caption{\label{tab:mcmc-constraints} 68\% confidence level marginalized cosmological constraints in \LCDM$ $ and wCDM for $\redmagic$ and the optimal $\maglim$ and flux-limited samples. }
	\begin{tabular}{lcccc}
		\hline
		\hline
		 \textsc{Lens sample}  & $\sigma(\Omega_m)$  & $\sigma(\sigma_8)$ & $\sigma(S_8)$ & $\sigma(w)$ \Tstrut \\\hline
		
		&\multicolumn{4}{c}{\textbf{\LCDM}}\Tstrut \\ 
		
		$\redmagic$ 			& 0.019 & 0.043 & 0.022 & --\Tstrut \\
		$\maglim$ 				 & 0.014  & 0.035	& 0.019 & -- \\
		\textsc{Flux-limited} &  0.017 & 0.037 & 0.018 & --\Bstrut\\\hline
		
		& \multicolumn{4}{c}{\textbf{wCDM}}\Tstrut \\\ 
		$\redmagic$ 			& 0.031 & 0.048 & 0.031 & 0.20\Tstrut \\
       $\maglim$ 				& 0.028  & 0.040	& 0.027 & 0.18 \\
       \textsc{Flux-limited} & 0.029  & 0.043 &  0.024 & 0.19\Bstrut\\\hline
		
	\end{tabular}
\end{table}

We then fix $w$ and compare the constraints on $\Omega_m$, $\sigma_8$ and $S_8$ assuming a \LCDM$ $ cosmological model. In Figure~\ref{fig:lcdm-mcmc} and Table~\ref{tab:mcmc-constraints} we show the constraints on these parameters from the combination of galaxy clustering and galaxy-galaxy lensing. In this case, we find a greater difference in the constraining power of the two samples. In particular, while the increase on
$\sigma_8$ and $S_8$ with $\maglim$ with respect to \redmagic is similar (around 19\% and 11\%, respectively), the constraints on $\Omega_m$ show a $27\%$ improvement compared to \redmagic. Thus, it seems that most of the gain in constraining power on $w$ in Figure~\ref{fig:wcdm-mcmc} has now been absorbed by $\Omega_m$.
Similarly to the wCDM case, the flux-limited sample yields worse constraints on $\Omega_m$ and $\sigma_8$ with respect to $\maglim$, with a difference of 18\% and 6\%, respectively, while it improves the constraints on $S_8$ compared to $\maglim$ by 6\% .

As discussed before, the fact we find tighter cosmological constraints with $\maglim$ is evidently due to the greater number of galaxies (2-3 times higher) and increased depth compared to \redmagic, reaching $z=1.05$ instead of $z=0.95$. If we included the shear 2-point correlation functions in our data vector, i.e. if we considered a 3$\times$2pt analysis, the difference between the two lens samples would be lower because the constraints would be dominated by the cosmic shear signal. However, this increase in depth of the $\maglim $ sample would be particularly advantageous when combining the 3x2pt analysis with CMB lensing (5$\times$2pt, see \cite{DESY1_5x2}), as the $\maglim $ sample will have a greater overlap with the CMB lensing kernel, providing a higher signal-to-noise ratio of galaxy clustering and CMB lensing cross-correlations.

\begin{figure}
	\begin{center}
		\includegraphics[width=\linewidth]{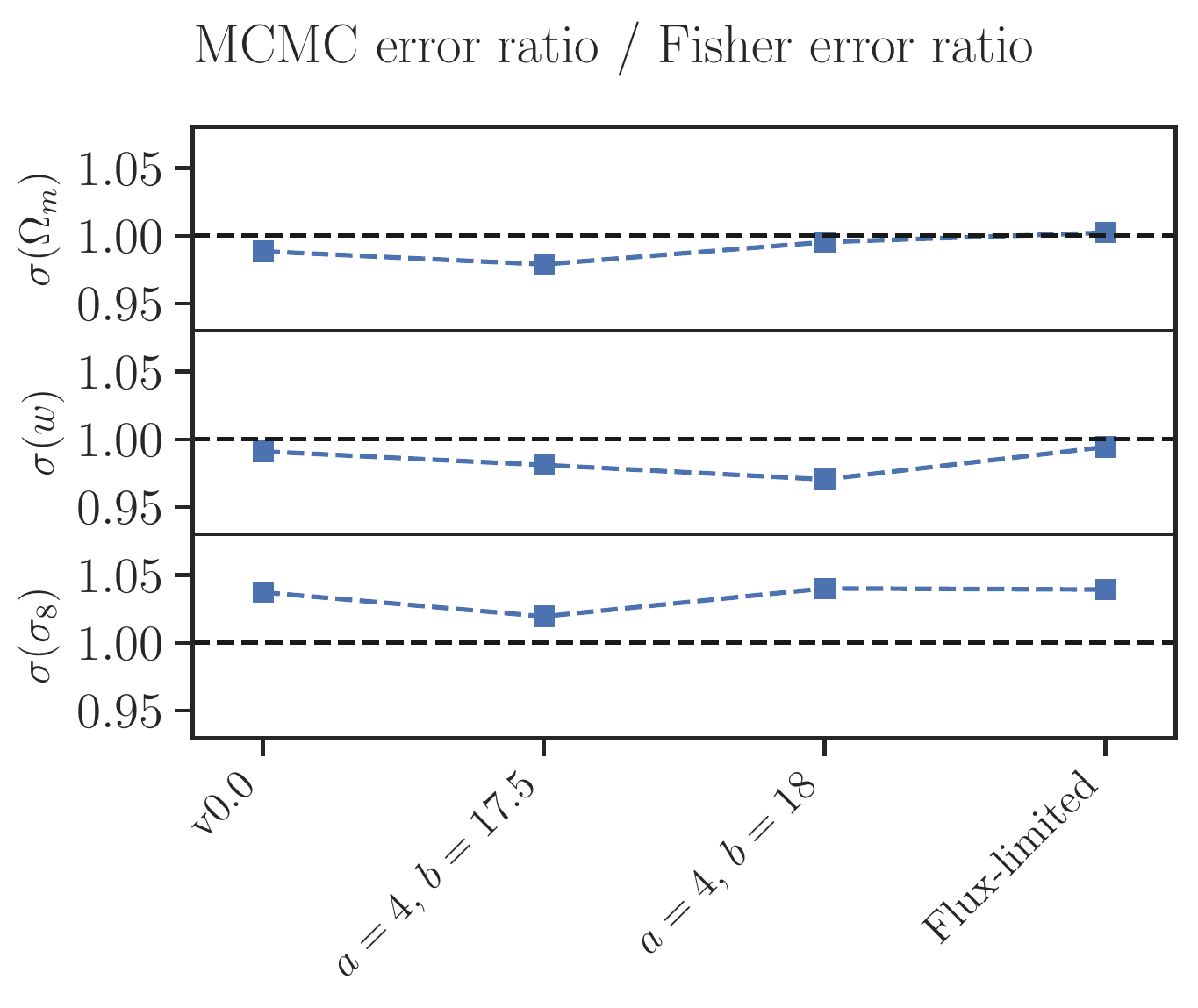}
		\caption{
				For each parameter $i$ and sample $j$, error ratio with respect to $\redmagic$ from MCMC divided by the equivalent error ratio from Fisher forecasts, i.e. $[\sigma_j^i/\sigma_{\textsc{red}}^i]_{\mathrm{MCMC}}/[\sigma_j^i/\sigma_{\textsc{red}}^i]_{\mathrm{Fisher}}$.  The samples considered are the flux-limited sample and a few definitions of the $\maglim$ sample, including the optimal one ($a=4,\, b=18$). Values larger (lower) than 1 indicate that the Fisher forecasts overestimate (underestimate) the gains of that sample with respect to $\redmagic$ compared to MCMC.  }
		\label{fig:mcmc-fisher-ratio}
	\end{center}
\end{figure}

Now that we have MCMC constraints for the different samples, in Figure~\ref{fig:mcmc-fisher-ratio} we turn back to compare the gains with respect to \redmagic with the results obtained with Fisher matrices in Sec.~\ref{sec:sampleoptimization}. For each parameter $i$ and sample $j$, we divide the error ratios obtained from MCMC with the error ratios using Fisher, $[\sigma_j^i/\sigma_{\textsc{red}}^i]_{\mathrm{MCMC}}/[\sigma_j^i/\sigma_{\textsc{red}}^i]_{\mathrm{Fisher}}$, where $\sigma_{\textsc{red}}$ denotes the constraints for \redmagic. In this way we can assess the level of uncertainty in our conclusions when using Fisher matrices, despite the offset with respect to MCMC constraints due to the non-gaussianity of the posteriors. The dashed line in Figure~\ref{fig:mcmc-fisher-ratio} denotes perfect agreement between MCMC and Fisher in the $\sigma$ errors when normalizing by \redmagic, while values larger (lower) than 1 indicate that  Fisher overestimates (underestimates) the gains. Figure~\ref{fig:mcmc-fisher-ratio} shows that the difference between MCMC and Fisher error ratios is less than 5\%. However, the variance of this difference across samples is small, having a scatter of $1-2\%$, in general. This is actually the level of impact in our conclusions when using Fisher, since in this paper we generally compare the gains of two different samples normalized by \redmagic.

\section{Sensitivity to Analysis Choices}
\label{sec:analysis-choices}

\subsection{Tomographic binning and cross-correlations}
\label{sec:tomographic-binning}

In this section we test the impact of the choice of tomographic binning of the $\maglim $ sample and the inclusion of galaxy clustering cross-correlations between redshift bins. We run $2\times2$pt Fisher forecasts for each of the tomographic-bin cases considered and compare the constraints on $\Omega_m$, $\sigma_8$, and $w$.

\begin{figure}
\begin{center}
\includegraphics[width=\linewidth]{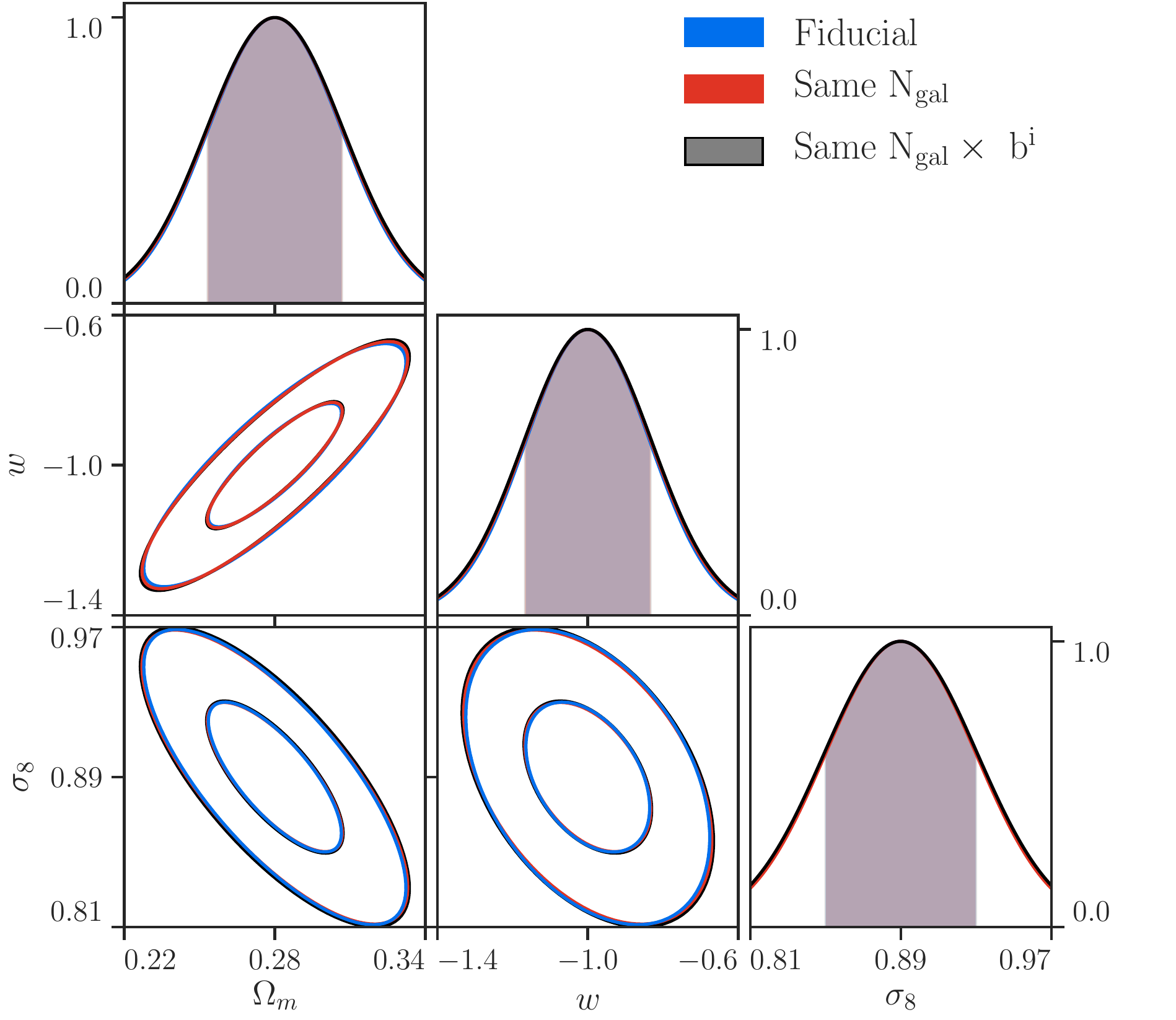}
\caption{Comparison of 2$\times$2pt Fisher constraints considering different tomographic binnings for the DES Y3 $\maglim $ sample, as described in Table~\ref{tab: maglim z binning 1}. For the sources we use the DES Y1 \metacal sample.}
\label{fig: maglim zbinning}
\end{center}
\end{figure}

Throughout this section we maintain the same global $z$ range as the fiducial sample, i.e. $0.2 < z < 1.05$. We first vary the edges of the tomographic binning, putting together two new configurations in which we balance the number of galaxies weighted by the galaxy bias in each redshift bin, `\emph{same $N_{\mathrm{gal}}\times b^i$}', and `\emph{same $N_{\mathrm{gal}}$}'. The galaxy bias values we consider are listed in Table~\ref{tab: priors}, and the definition of these $z$ binnings is shown in Table~\ref{tab: maglim z binning 1}. The motivation for balancing the number of galaxies (weighted by the galaxy bias) is to have a more uniform signal-to-noise ratio across redshift, as the shot noise $\propto 1/N_{\mathrm{gal}}$ and the signal is proportional to the bias (see Eq.~\eqref{eq:density kernel}). However, as we can see in Figure~\ref{fig: maglim zbinning}, where we compare the constraints coming from these different tomographic binnings, our choice of binning does not appreciably impact the 2$\times$2pt cosmological constraints.

\begin{table}
	\centering
	 	\caption{\label{tab: maglim z binning 1} Different tomographic binning configurations for the $\maglim $ sample, considering variations in the edges of the $z$ bins.}
		\begin{tabular}{ccc}
			\hline \hline
			\textsc{ Fiducial}  &  \textsc{ Same $N_{\mathrm{gal}}$ } &  \textsc{ Same $N_{\mathrm{gal}}\times b^i$ } \\
			\hline
			0.20 -- 0.35 & 0.20 -- 0.36 & 0.20 -- 0.40 \Tstrut \\
			0.35 -- 0.50 & 0.36 -- 0.52 & 0.40 -- 0.55 \\
			0.50 -- 0.65 & 0.52 -- 0.69 & 0.55 -- 0.72 \\
			0.65 -- 0.80 & 0.69 -- 0.82 & 0.72 -- 0.85 \\
			0.80 -- 0.95 & 0.82 -- 0.93 & 0.85 -- 0.95 \\
			0.95 -- 1.05 & 0.93 -- 1.05 &  0.95 -- 1.05 \Bstrut   \\  \hline
		\end{tabular}
\end{table}

\begin{table}
	\caption{\label{tab: maglim z binning 2} Different tomographic binning configurations for the $\maglim $ sample, considering variations in the number of $z$ bins. The case with 6 $z$ bins corresponds to the fiducial tomographic binning (see e.g. Table~\ref{tab: maglim z binning 1}).}
	\begin{tabular}{cccc}
		\hline \hline
		\textsc{4 $z$-bins } &   \textsc{5 $z$-bins }&  \textsc{7 $z$-bins } &   \textsc{8 $z$-bins }\\
		\hline
		0.20 -- 0.44 & 0.20 -- 0.40 & 0.20 -- 0.35 & 0.20 -- 0.31\Tstrut \\
		0.44 -- 0.69 & 0.40 -- 0.60 & 0.35 -- 0.50 & 0.31 -- 0.44 \\
		0.69 -- 0.87 & 0.60 -- 0.77 & 0.50 -- 0.64 & 0.44 -- 0.57 \\
		0.87 -- 1.05 & 0.77 -- 0.90  & 0.64 -- 0.77 & 0.57 -- 0.69\\
		& 0.90 -- 1.05 & 0.77 -- 0.86 & 0.69 -- 0.79 \\
		& 					 & 0.86 -- 0.95 & 0.79 -- 0.87 \\
		&					 & 0.95 -- 1.05 & 0.87 -- 0.96 \\
		&					&						& 0.96 -- 1.05 \Bstrut   \\  \hline
	\end{tabular}
\end{table}

\begin{figure}
	\begin{center}
		\includegraphics[width=\linewidth]{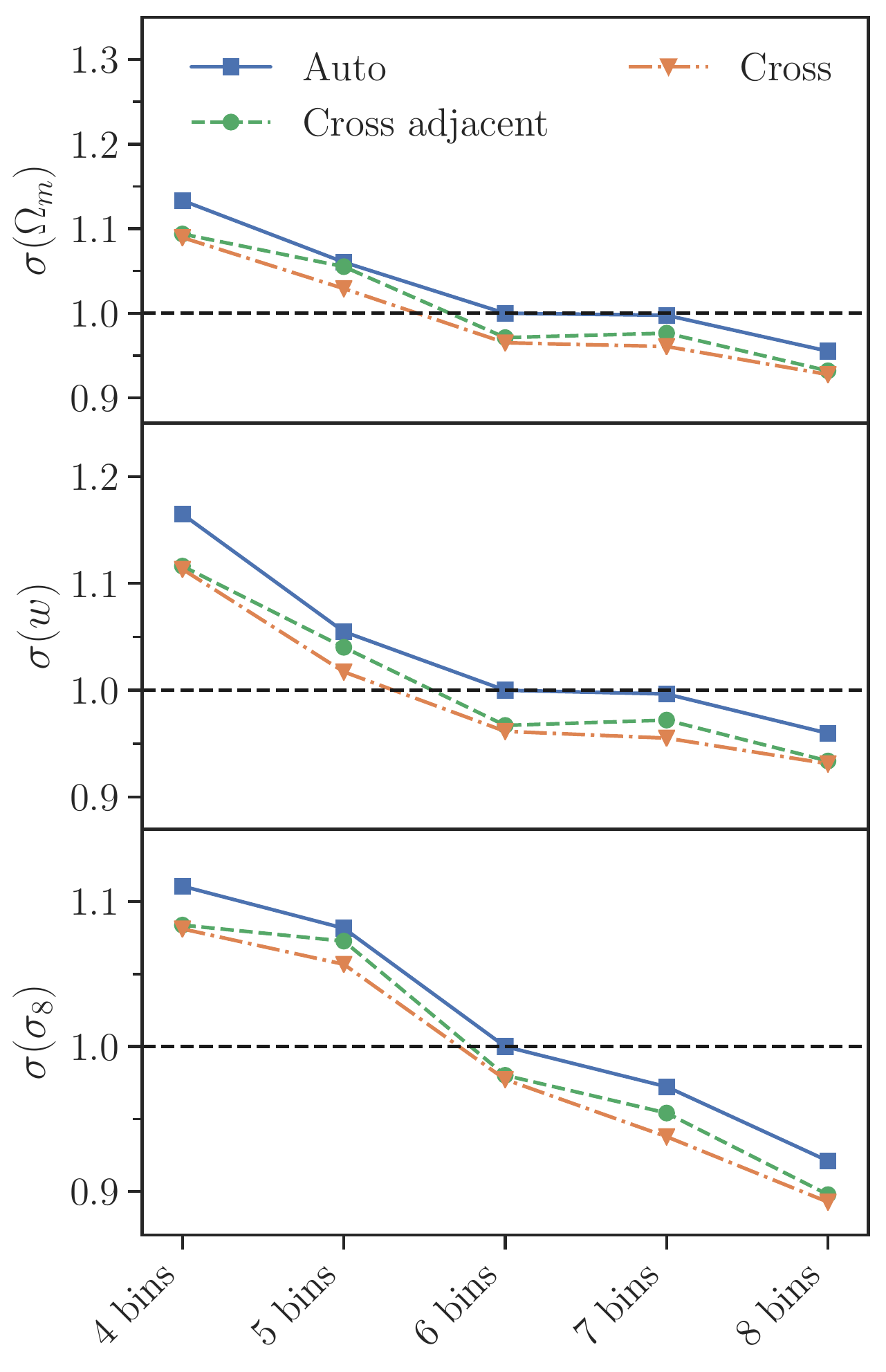}
		\caption{Comparison of 2$\times$2pt Fisher constraints considering different number of tomographic bins for the DES Y3 $\maglim $ sample, as described in Table~\ref{tab: maglim z binning 2}. For the sources we use the DES Y1 \metacal sample. All constraints are normalized by the fiducial  ('auto' with 6 redshift bins). We compare the gains obtained when only galaxy clustering auto-correlations (solid blue) are included, with the cases in which we also include cross-correlations with adjacent  tomographic bins (dashed green), and when all  cross- correlations among tomographic bins are included (dash-dotted orange). }
		\label{fig: maglim num bins cross}
	\end{center}
\end{figure}

We then vary the number of tomographic bins in which we divide the sample in the range $0.2 < z < 1.05$. Our fiducial tomographic binning consists of 6 $z$ bins, and we consider additionally sample selections  split in 4, 5, 7, and 8 $z$ bins. See Table~\ref{tab: maglim z binning 2}  for the details of the $z$ binning for each one of these cases. 
In Figure~\ref{fig: maglim num bins cross} we compare the estimated Fisher $2\times2$pt constraints from each one of these sample selections with our fiducial choice of 6 $z$ bins, and we examine the importance of including galaxy clustering cross-correlations between redshift bins. The motivation for the latter is that, as seen in Figure~\ref{fig:nz-optimal-samples}, the $\maglim $ sample has more overlap between $z$ bins than the \redmagic sample, so galaxy clustering cross-correlations could become important for our analysis. In addition, \cite{Tanoglidis2019} shows that the improvement on the $\Omega_{\rm m}$ and $\sigma_8$ constraints can be greatly increased with the number of $z$ bins and the inclusion of cross-correlations between $z$ bins, especially for samples with large overlap between bins.

 We can draw several conclusions from Figure~\ref{fig: maglim num bins cross} (see Table~\ref{tab:secvi-summary} for a quantitative summary of the most relevant cases). First, we find that \emph{reducing} the number of $z$ bins degrades the cosmological constraints. This makes sense, as reducing the number of bins while keeping fixed the total $z$ range to be covered effectively increases the width of the redshift distributions and, as shown in \cite{Asorey2012}, there is a loss of information when projecting the 3D power spectrum into angular tomographic bins, with that loss being larger the wider the redshift bins. This is due to the fact that broad bins average down radial power on scales smaller than the bin width. More concretely, when splitting the sample in 4 tomographic bins instead of 6, the constraints degrade up to 13\% on $\Omega_m$, 16\% on $w$ and 11\% on $\sigma_8$.
 
Second, \emph{increasing} the number of redshift bins improves the constraints, but the impact is smaller. Only when we split the sample in 8 $z$ bins do we start to find some significant improvement in all three parameters, and especially on $\sigma_8$. In particular, in this case the constraints improve by 4\% on $\Omega_m$ and $w$, and 8\% on $\sigma_8$, with respect to the fiducial. 
In spite of this, we keep the 6 $z$ bins tomographic binning as our fiducial, considering that splitting into a larger number of tomographic bins would require a better understanding of the tails of the redshift distributions, which is likely not captured by our treatment of \photoz uncertainties (just a shift to the mean of the distribution). Another motivation for not splitting into a larger number of bins is to avoid  numerical instabilities in the computation of the analytical non-Gaussian covariance. We note that the results from Figure~\ref{fig: maglim num bins cross}  may change slightly with the inclusion of non-gaussian terms in the covariance. That is due to the non-gaussian terms being unaffected by the change in number densities in each tomographic bin, while the Gaussian part of the covariance does vary with the number densities.

Last, we study the impact of including galaxy clustering cross-correlations in our analysis. Ref.~\cite{Tanoglidis2019} shows that, for a flux-limited sample, the improvement on the cosmological constraints can be greatly increased with the number of $z$ bins and the inclusion of cross-correlations between $z$ bins.  In \cite{Tanoglidis2019}  the authors consider only galaxy clustering, and fix all parameters except for $\Omega_m$, $\sigma_8$, and the photo-$z$ nuisance parameters. We have attempted to reproduce their results, and while we do not find the same level of gains on the constraints, we observe the same tendency.  In Figure~\ref{fig: maglim num bins cross} we repeat this study for the $\maglim $ sample, but now varying all parameters listed in Table~\ref{tab: priors}, and including galaxy-galaxy lensing.  We find that there is not much improvement to be gained with the inclusion of galaxy clustering cross-correlations between $z$ bins (a $3-4\%$ gain in the three cosmological parameters), and that this relative gain does not depend on the number of tomographic bins considered. We also explore the possibility of including only galaxy clustering cross-correlations with adjacent $z$ bins, which is where the overlap between bins is the largest, finding in general very similar constraints compared to when we include all cross-correlations between $z$ bins.

\begin{figure}
	\begin{center}
		\includegraphics[width=\linewidth]{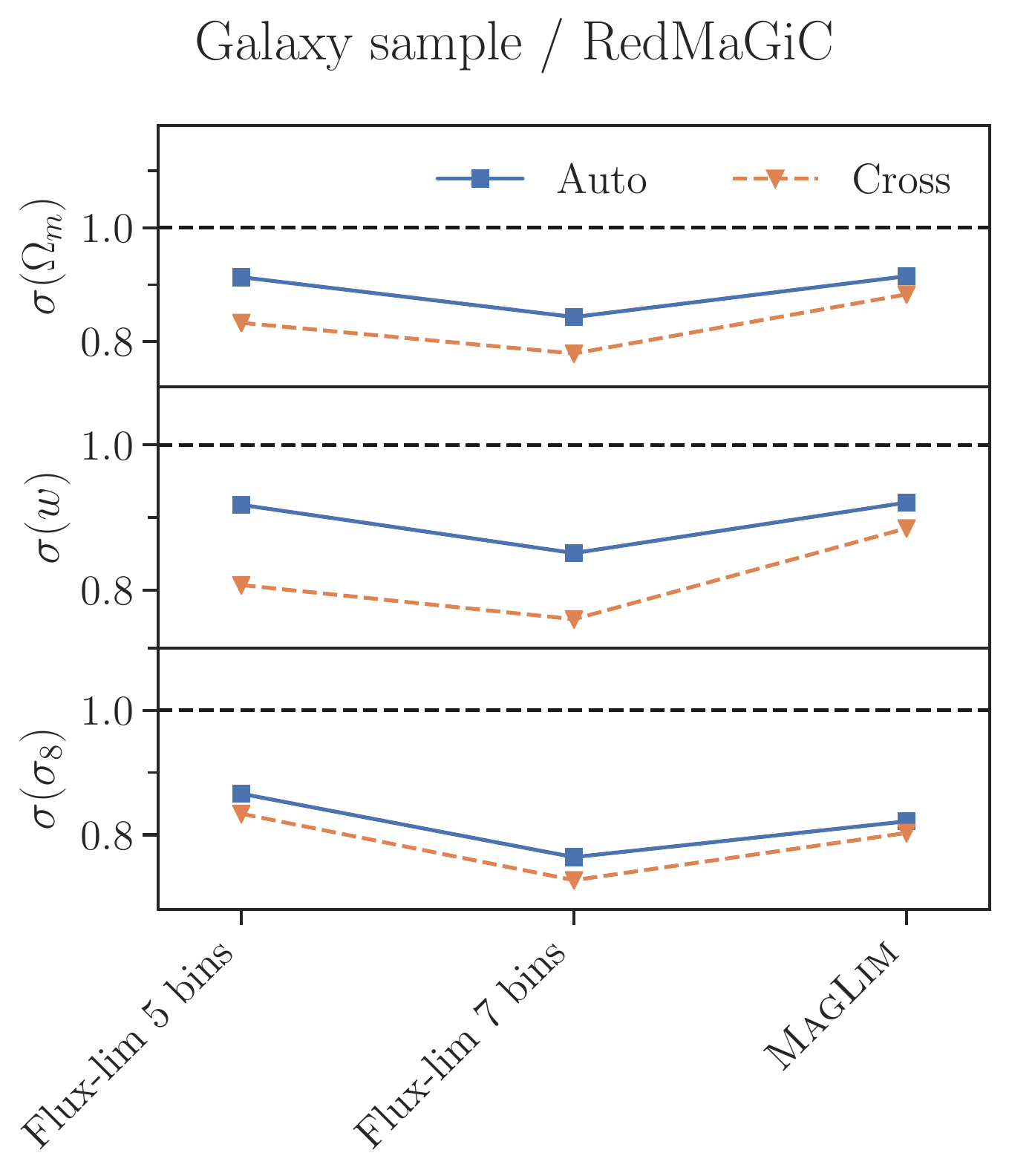}
		\caption{Comparison of 2$\times$2pt standard deviations on $\Omega_m$, $w$, and $\sigma_8$ for different numbers of tomographic bins for the flux-limited sample compared to the fiducial $\maglim$ sample. All constraints are normalized by the  $\redmagic$ estimates. We explore the potential gains when including all galaxy clustering cross-correlations among tomographic bins (dashed orange) with the baseline approach, i. e. including only galaxy clustering auto-correlations (solid blue). All three lens samples are built from DES Y3 data, while the sources are from the DES Y1 \metacal sample.}
		\label{fig: flux lim num bins cross}
	\end{center}
\end{figure}

We also explore the potential gains on the 2$\times$2pt constraints from the flux-limited sample when including all cross-correlations between $z$ bins and splitting the sample in a larger number of bins than the fiducial (5 bins). In particular, we divide the sample in 7 tomographic bins in these redshift ranges, aiming for a balanced number density across bins: [0.2, 0.35, 0.45, 0.55, 0.65, 0.75, 0.85, 1.05]. In Figure~\ref{fig: flux lim num bins cross}  we observe that the gain on the constraints from the inclusion of cross-correlations is larger than for $\maglim$, as expected, since the flux-limited sample has broader redshift distributions (see Figure~\ref{fig:nz-optimal-samples}). In particular, with this sample, including cross-correlations improves the constraints by $\sim8\%$ on $\Omega_m$, $\sim11\%$ on $w$, and $\sim4\%$ on $\sigma_8$. 
Similarly, increasing the number of tomographic bins improves the constraints by a larger amount compared to $\maglim$. When splitting the sample in 7 tomographic bins, the gains on the cosmological parameters with respect to the fiducial (5 bins) are of $7-8\%$ on $\Omega_m$ and $w$, and $12\%$ on $\sigma_8$.
Thus, by splitting the flux-limited sample in a larger number of bins we can already obtain tighter cosmological constraints than \maglim in all three parameters. More concretely, dividing the sample in 7 tomographic bins yields constraints tighter than the fiducial \maglim (with 6 bins) by $7-8\%$. 

In practice there are several complications in considering a large number of bins and in the inclusion of  cross-correlations, the main ones being a much more stringent requirement for the control of the tails of the redshift distributions and a larger covariance.
Moreover, as mentioned before, it is not clear that a \photoz shift parameter is enough to account for these uncertainties.  For these reasons, we will focus on the \maglim sample in follow-up work with DES Y3 data. Nonetheless, flux-limited samples are promising and worth exploring in future studies. 


\subsection{Galaxy bias}
\label{sec:galaxy-bias}

Throughout this work we assume certain fiducial values for the galaxy bias for each one of the samples (see Table~\ref{tab: priors}) to generate the theory data vectors, which we do not vary when considering different sample definitions for the $\maglim $ and flux-limited samples in Sec.~\ref{sec:sampleoptimization}. In order to test the dependency of the constraints on the fiducial galaxy bias assumed, we run Fisher forecasts with completely different galaxy bias values and compare the 2$\times$2pt constraints on $\Omega_m$, $w$, and $\sigma_8$, finding almost no difference in our results. In particular, for the $\maglim $ sample we run a forecast assuming a constant value of $2.0$ for the galaxy bias in all redshift bins, finding a difference in the constraints on the cosmological parameters of $\sim 1\%$ or less.

For the flux-limited sample, following \cite{Tanoglidis2019} we run a forecast assuming a galaxy bias that evolves as $b(z)=1+z$, hence the galaxy bias in each redshift bin is $b^i = 1 +\bar{z}^i$, with $\bar{z}^i$ being the mean redshift in that tomographic bin. For this sample, that corresponds to $b=[1.33, 1.46, 1.56, 1.72, 1.88]$. Compared to assuming the fiducial values in Table~\ref{tab: priors}, the resulting constraints on $\Omega_m$, $w$ and $\sigma_8$ differ by less than $\sim 0.5\%$. Thus the conclusions from this work do not depend on the galaxy bias assumed.

\subsection{Photometric redshift uncertainties}
\label{sec:photoz-uncertainties}

As described in Sec.~\ref{sec:params-and-priors}, we quantify the uncertainties in the redshift distributions by introducing a \photoz shift parameter in each redshift bin, $\Delta z^i$, that we marginalize over in our analysis assuming a Gaussian prior with a certain $\sigma^i$. 
In this section we test the dependency of the $\maglim$ and flux-limited sample gains on the width $\sigma$ of the prior assumed. For this purpose, we investigate a pessimistic scenario for \maglim in which the Gaussian priors on $\Delta z^i$ are two times wider than the fiducial in Table~\ref{tab: priors}. We find that in this case the constraints degrade by  $6-7\%$ for $\Omega_m$ and $w$, and about $3\%$ for $\sigma_8$.

 Similarly, for the flux-limited sample we also test the impact on the constraints when increasing the width of the priors. In particular, we run a Fisher forecast with Gaussian priors three times as wide as the \maglim ones (the fiducial priors are two times wider, see Table~\ref{tab: priors}). The resulting constraints are degraded by $5-8\%$  compared to the fiducial \photoz prior for the flux-limited sample.

\subsection{Weak lensing systematics}

 In all our analyses we assume the DES Y1 source sample from \cite{DESY1_3x2}, and its corresponding priors for the weak lensing related nuisance parameters: the shear calibration bias in each source redshift bin, $m^i$, the intrinsic alignment parameters ($A_{\mathrm{IA}}$, and $\alpha_{\mathrm{IA}}$), and the source photo-z shift parameters in each bin, $\Delta z _{s}^i$.  We expect some improvement in our control of these systematics for the upcoming DES Y3 and Y6 analyses that will tighten the priors on these nuisance parameters. In this section we investigate to what extent our forecasts are limited by our (prior) knowledge of the weak lensing systematics. For this purpose we consider the ideal scenario in which we perfectly know the values of these systematic parameters, i.e. we fix them in our analysis. We find that, for the $\maglim $ sample, we can improve the constraints up to $\sim10\%$ for $\Omega_m$ and $w$, and $15\%$ for $\sigma_8$. The constraints on \redmagic also improve in a similar manner, nonetheless the larger number density of $\maglim $ could be more important in this scenario in which the weak lensing systematics are not a bottleneck. Comparing the 2$\times2$pt constraints with fixed weak lensing systematics from $\maglim $ and \redmagic, we find that the relative constraining power of the former remains similar for $\Omega_m$ and $w$ and improves by $3\%$ for $\sigma_8$ with respect to what we obtain for the two samples when marginalizing the weak lensing nuisance parameters. 


The gain in constraining power that the $\maglim $ sample offers compared to \redmagic is mainly due to the larger number density, as that reduces the shot noise contribution in the covariance. We also explore in this section how much are we limited by the shot noise of the lens sample. We compute a covariance matrix setting the galaxy clustering shot noise contribution to zero (equivalent to assuming a practically infinite number density), and we find for the $\maglim $ sample an improvement of $6\%$ for $\Omega_m$, $3\%$ for $w$, and $12\%$ for $\sigma_8$ with respect to the fiducial case. Therefore, the $\maglim $ sample is relatively close to the limit without shot noise.

\begin{table}
	\centering
	\caption{\label{tab:secvi-summary} Percentage gains in the $\Omega_m$, $\sigma_8$ and $w$ standard deviations with respect to the fiducial for $\maglim$, considering the most relevant alternative analysis choices discussed in Sec.~\ref{sec:analysis-choices}. Negative values indicate a decrease in the constraining power compared to the fiducial.}
	\begin{tabular}{lrrr}
		\hline
		\hline
  & $\sigma(\Omega_m)$  & $\sigma(\sigma_8)$ & $\sigma(w)$ \Tstrut \\\hline

	auto $+$ cross 				  &  3.5\% & 2.3\% & 3.8\%\Tstrut \\
	4 tomographic bins 			& -13.3\% & -11.1\% & -16.5\%\\
	8 tomographic bins 		    & 4.5\%  & 7.9\% & 4.0\%  \\
	x2 photo-z shift priors     & -6.3\%& -3.0\% & -6.6\% \\
	fixed WL systematics 		&  9.8\% & 15.1\% & 9.4\% \\
	infinite number density 	&  5.7\% & 11.8\% & 2.8\%\Bstrut\\\hline
		
	\end{tabular}
\end{table}

\section{Conclusions}
\label{sec:conclusions}

In this work we define an optimized lens sample for DES Y3 that serves as an alternative to \redmagic for cosmological analyses involving galaxy clustering measurements.  Assuming the DES Y1 $\metacal$ sample for the sources, we compare  the cosmological constraints on $\Omega_m$, $\sigma_8$ and $w$ from the joint analysis of galaxy clustering and galaxy-galaxy lensing for different lens sample definitions.  The main conclusions that we obtain are:

\begin{enumerate}[label=(\roman*)]
	\item We explore which flux-limited samples are optimal in terms of their cosmological constraints. We consider, first, samples with a magnitude cut in the $i$ band  depending linearly with redshift and, second, samples defined with an overall limiting magnitude. We see that many of the samples considered yield constraints similar to or better than \redmagic due to the superior number density. We find that the optimal sample overall, dubbed $\maglim $, is defined with $i<4 z_{\rm phot} +18$, and that it improves the figure of merit of the pair combinations of $\Omega_m$, $w$, and $\sigma_8$ by 40\% with respect to \redmagic (see Figure~\ref{fig: fisher_comp}). 
	\item 
	$\maglim $ has between 2 and 3 times more galaxies than \redmagic while having $\sim30\%$ wider redshift distributions. 
	We compare the cosmological constraints from 2$\times$2pt MCMC simulated likelihood analyses, after marginalizing over the same set of $\sim 20$ cosmological and nuisance parameters as in the DES Y1 analysis \cite{DESY1_3x2}, finding that the $\maglim $ sample provides 10\% tighter constraints on $\Omega_m$,  12-13\% on $w$ and $S_8$, and 16\% on $\sigma_8$  with respect to \redmagic in $w$CDM. We then consider a \LCDM$ $ scenario, fixing $w=-1$, finding improvements on  $\Omega_m$ of 27 \% compared to \redmagic, while the gains on  $\sigma_8$ and $S_8$ are respectively 19\% and 11\%. 

	\item We study how the performance of the optimized sample varies for different analysis choices, which we summarize in Table~\ref{tab:secvi-summary}. We find that changing the galaxy bias and the tomographic binning (given a fixed number of redshift bins)  does not impact the 2$\times$2pt constraints. In turn, reducing the number of bins degrades the constraints, and increasing it improves them slightly (by $4-8\%$). Independently of the number of bins considered, we find that there is little to be gained with the inclusion of galaxy clustering cross-correlations.  We also test the impact of changing the width of the priors on the \photoz shift parameters.  In a pessimistic scenario, with priors twice as big for $\maglim$, the constraints degrade by $\sim6-7\%$ for $\Omega_m$ and $w$, and about 3\% for $\sigma_8$. 
	Last, we find 
		that  $\maglim $ is relatively close to the sample variance limited regime. If we assume an infinite number density in the covariance, the constraints improve by 6\% for $\Omega_m$, 3\% for $w$, and  12\% for $\sigma_8$ with respect to the fiducial. 

\item For flux-limited samples with a flat magnitude cut, the optimization leads to a limiting magnitude of $i<22.2$.  This has one order of magnitude more galaxies per redshift bin compared to $\maglim $, with $\sim20\%$ wider redshift distributions. Although this sample provides tighter constraints than \redmagic, it is slightly less constraining than $\maglim $. If we divide the sample in a large number of tomographic bins, we obtain constraints tighter than $\maglim$ by $7-8\%$. Including galaxy clustering cross-correlations can further improve the constraints by $5-10\%$. In this limit however one probably needs to include further nuisance parameters and a realistic analysis becomes more complex. 


\item Although not discussed in detail, \maglim does lead to a higher signal-to-noise ratio of galaxy clustering and CMB lensing cross-correlations due to its increased redshift reach compared to \redmagic. This translates into a larger forecasted constraining power for this probe in DES Y3 (see \cite{DESY1_5x2} for the Y1 equivalent).

\end{enumerate}


The results presented in this paper have been derived using a likelihood setup as realistic as possible, matching the one in DES Y1. We have already confirmed that our results are robust with respect to the addition of the main characteristics of a Y3 analysis, like the source samples and other effects such as the inclusion of non-Limber modeling \cite{2020JCAP...05..010F}, point-mass marginalization \cite{2020MNRAS.491.5498M}, or non-Gaussian covariances \cite{2020arXiv200404833F}. Moreover, we have validated in \citep[in prep.]{DESY3maglim} the scale cuts for linear/non-linear bias in a Y3 analysis, finding that we can use the same scale cuts for both \maglim and \redmagic, as assumed in this work.   
  However there are a number of assumptions that will need to be re-evaluated in an actual data analysis, most notably the exact treatment of lens redshift distributions and their associated errors. Despite this, using a $\maglim $ type of sample for the cosmological analysis in DES Y3 (or similar datasets) is promising, both (i) to yield competitive or tighter 3$\times$2pt constraints than current standard lens samples and (ii)  to provide a robustness test for the dependence of these constraints with the  foreground (lens) sample. Such an analysis will also open the window to defining optimal and well calibrated samples for different probes. Last, flux-limited samples with a simple selection, such as \maglim, are likely to be easily reproducible in simulations and to have a more straightforward HOD modeling on small scales, where the reduced shot noise of this kind of sample would be particularly beneficial. Addressing the required steps for a cosmological data analysis with \maglim will be the focus of follow-up work.



\begin{acknowledgements}

We thank David Weinberg for helpful suggestions and discussions.  

Funding for the DES Projects has been provided by the U.S. Department of Energy, the U.S. National Science Foundation, the Ministry of Science and Education of Spain, 
the Science and Technology Facilities Council of the United Kingdom, the Higher Education Funding Council for England, the National Center for Supercomputing 
Applications at the University of Illinois at Urbana-Champaign, the Kavli Institute of Cosmological Physics at the University of Chicago, 
the Center for Cosmology and Astro-Particle Physics at the Ohio State University,
the Mitchell Institute for Fundamental Physics and Astronomy at Texas A\&M University, Financiadora de Estudos e Projetos, 
Funda{\c c}{\~a}o Carlos Chagas Filho de Amparo {\`a} Pesquisa do Estado do Rio de Janeiro, Conselho Nacional de Desenvolvimento Cient{\'i}fico e Tecnol{\'o}gico and 
the Minist{\'e}rio da Ci{\^e}ncia, Tecnologia e Inova{\c c}{\~a}o, the Deutsche Forschungsgemeinschaft and the Collaborating Institutions in the Dark Energy Survey. 

The Collaborating Institutions are Argonne National Laboratory, the University of California at Santa Cruz, the University of Cambridge, Centro de Investigaciones Energ{\'e}ticas, 
Medioambientales y Tecnol{\'o}gicas-Madrid, the University of Chicago, University College London, the DES-Brazil Consortium, the University of Edinburgh, 
the Eidgen{\"o}ssische Technische Hochschule (ETH) Z{\"u}rich, 
Fermi National Accelerator Laboratory, the University of Illinois at Urbana-Champaign, the Institut de Ci{\`e}ncies de l'Espai (IEEC/CSIC), 
the Institut de F{\'i}sica d'Altes Energies, Lawrence Berkeley National Laboratory, the Ludwig-Maximilians Universit{\"a}t M{\"u}nchen and the associated Excellence Cluster Universe, 
the University of Michigan, NFS's NOIRLab, the University of Nottingham, The Ohio State University, the University of Pennsylvania, the University of Portsmouth, 
SLAC National Accelerator Laboratory, Stanford University, the University of Sussex, Texas A\&M University, and the OzDES Membership Consortium.

Based in part on observations at Cerro Tololo Inter-American Observatory at NSF's NOIRLab (NOIRLab Prop. ID 2012B-0001; PI: J. Frieman), which is managed by the Association of Universities for Research in Astronomy (AURA) under a cooperative agreement with the National Science Foundation.

The DES data management system is supported by the National Science Foundation under Grant Numbers AST-1138766 and AST-1536171.
The DES participants from Spanish institutions are partially supported by MICINN under grants ESP2017-89838, PGC2018-094773, PGC2018-102021, SEV-2016-0588, SEV-2016-0597, and MDM-2015-0509, some of which include ERDF funds from the European Union. IFAE is partially funded by the CERCA program of the Generalitat de Catalunya.
Research leading to these results has received funding from the European Research
Council under the European Union's Seventh Framework Program (FP7/2007-2013) including ERC grant agreements 240672, 291329, and 306478.
We  acknowledge support from the Brazilian Instituto Nacional de Ci\^encia
e Tecnologia (INCT) do e-Universo (CNPq grant 465376/2014-2).

This manuscript has been authored by Fermi Research Alliance, LLC under Contract No. DE-AC02-07CH11359 with the U.S. Department of Energy, Office of Science, Office of High Energy Physics.

This research used resources of the Ohio Supercomputer Center (OSC) \cite{OSC} and of the National Energy Research Scientific Computing Center (NERSC), a U.S. Department of Energy Office of Science User Facility operated under Contract No. DE-AC02-05CH11231. 

We acknowledge the use of the \textsc{CosmicFish} \cite{Cosmicfish,CosmicfishB} and \textsc{ChainConsumer}  \cite{Chainconsumer} packages to plot the Fisher and MCMC contours, respectively.

\end{acknowledgements}

\section*{Appendix}
	\appendix
\label{appendix}

\section{$\cosmolike $ vs. $\cosmosis$}
\label{sec:cov-comp}

\begin{figure}
	\begin{center}
		\includegraphics[width=\linewidth]{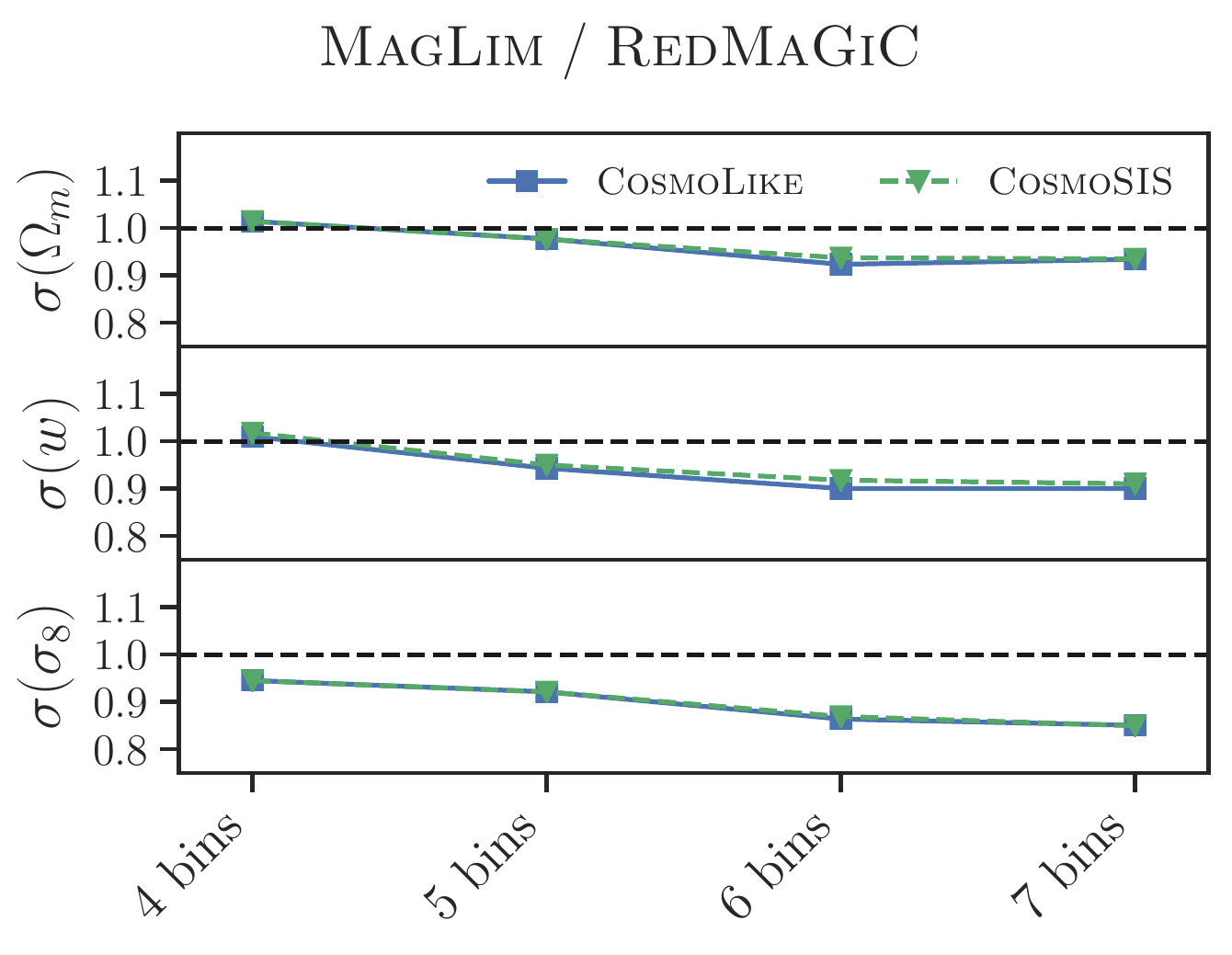}
		\caption{Standard deviations on $\Omega_m$, $w$ and $\sigma_8$ from different tomographic binnings of the $\maglim $ sample normalized by estimates from the \redmagic$ $ sample.  The constraints have been obtained using two different codes for the covariances: $\cosmolike $ and $\cosmosis$.  }
		\label{fig:cosmolike-cosmosis}
	\end{center}
\end{figure} 

	Throughout this work we use the  $\cosmolike $ and $\cosmosis $ codes interchangeably to compute the Gaussian analytical covariances we use for our forecasts. Here we compare the constraints obtained using covariances estimated from the two codes. In Figure~\ref{fig:cosmolike-cosmosis} we show the relative gain on $\Omega_m$, $w$ and $\sigma_8$ errors compared to $\redmagic $ for different tomographic binnings of the $\maglim $ sample (see Sec.~\ref{sec:tomographic-binning}). We compare the estimates using a covariance from $\cosmolike $ (solid blue) with those obtained using a covariance from $\cosmosis $  (dashed green), finding no difference in the constraints.


\bibliography{maglim.bib}

\end{document}